\documentclass[twocolumn,notitlepage,prx,superscriptaddress,longbibliography]{revtex4-1}
\usepackage{amsfonts}
\usepackage{textcomp}
\usepackage{times}
\usepackage{graphicx}
\usepackage{float}
\usepackage{latexsym,amsmath,amssymb,bm,euscript}
\usepackage{color}
\usepackage{subfigure}
\usepackage{epstopdf}
\usepackage[colorlinks=true,linkcolor=blue,citecolor=blue]{hyperref}
\usepackage{hyperref}
\usepackage{soul}
\usepackage[normalem]{ulem}
\usepackage{mathrsfs}
\usepackage{amsmath}
\usepackage{xspace}
\usepackage{textcomp}

\begin{document}

\title{Thermal Tensor Network Simulations of the Heisenberg Model on the
Bethe Lattice}
\author{Dai-Wei Qu}
\affiliation{Department of Physics, Key Laboratory of Micro-Nano Measurement-Manipulation
and Physics (Ministry of Education), Beihang University, Beijing 100191,
China}
\author{Wei Li}
\email{w.li@buaa.edu.cn}
\affiliation{Department of Physics, Key Laboratory of Micro-Nano Measurement-Manipulation
and Physics (Ministry of Education), Beihang University, Beijing 100191,
China}
\affiliation{International Research Institute of Multidisciplinary Science, Beihang
University, Beijing 100191, China}
\author{Tao Xiang}
\email{txiang@iphy.ac.cn}
\affiliation{Institute of Physics, Chinese Academy of Sciences, Beijing 100190, China}
\affiliation{School of Physics, University of Chinese Academy of Sciences, Beijing 100049, China}

\begin{abstract}
We have extended the canonical tree tensor network (TTN) method, which was
initially introduced to simulate the zero-temperature properties of quantum
lattice models on the Bethe lattice, to finite temperature simulations. By representing the thermal density matrix with a canonicalized tree tensor product
operator, we optimize the TTN and accurately evaluate the thermodynamic
quantities, including the internal energy, 
specific heat, and the spontaneous magnetization, etc, at various temperatures. By varying the anisotropic
coupling constant $\Delta$, we obtain the phase diagram of 
the spin-1/2 Heisenberg XXZ model on the Bethe lattice, 
where three kinds of magnetic ordered phases, namely the ferromagnetic, 
XY and antiferromagnetic ordered phases, are found in low temperatures 
and separated from the high-$T$ paramagnetic phase by a continuous thermal phase transition at $T_c$. 
The XY phase is separated from the other two phases by two
first-order phase transition lines at the symmetric coupling points $
\Delta=\pm 1$. We have also carried out a linear spin wave calculation
on the Bethe lattice, 
showing that the low-energy
magnetic excitations are always gapped, 
and find the 
obtained magnon gaps in very good agreement 
with those estimated from the TTN simulations.
Despite the gapped excitation spectrum, 
Goldstone-like transverse fluctuation modes, 
as a manifestation of spontaneous continuous symmetry breaking, 
are observed in the ordered magnetic phases with $|\Delta|\le 1$.
One remarkable feature there is that the prominent transverse correlation length reaches $\xi_c=1/\ln{(z-1)}$ for $T\leq T_c$, the maximal value allowed on a $z$-coordinated Bethe lattice. 
\end{abstract}

\maketitle
\section{Introduction}

\begin{figure}[!tbp]
\includegraphics[angle=0,width=1\linewidth]{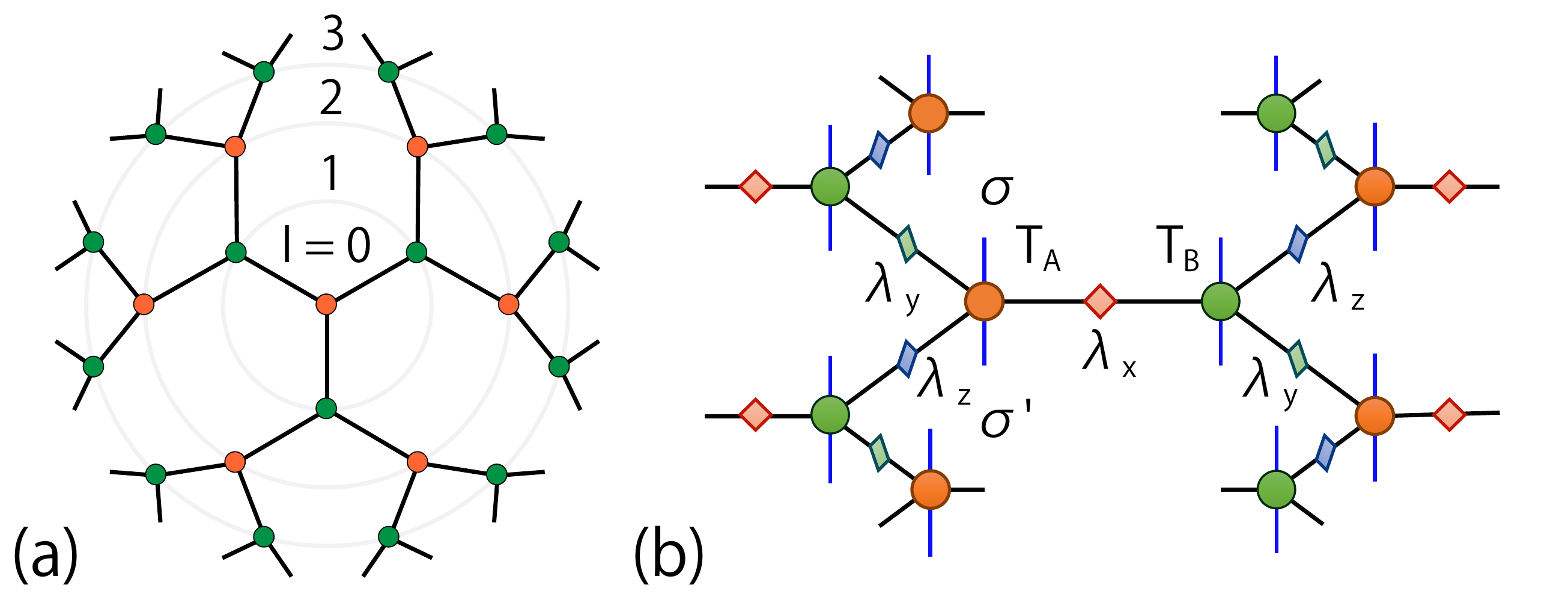}
\caption{(a) The $z=3$ Bethe lattice constitutes an infinite loop-free tree
structure. This lattice can be divided into two sublattices, labelled as $A$%
- (orange circles) and $B$-sublattice (green), respectively. Given a
``root'' site, the lattice sites on the Bethe lattice can be labelled
by the shell number $l=0,1,2,..., \infty$ they reside. (b) A TTN
representation of the density matrix, which contains two types of tensors,
namely the site tensors, $T_A$ and $T_B$, defined on the lattice nodes. 
The bond matrices, $\protect\lambda_x$, $\protect\lambda_y$, and $\protect%
\lambda_z$, are defined on the corresponding geometric bonds.}
\label{Fig:Tree}
\end{figure}

The Bethe approximation, a renowned cluster mean-field approach proposed by
Bethe in 1935 \cite{Bethe1935}, has played an important role in the studies of
cooperative phenomena and phase transitions of classical statistical models \cite
{Pathria2011}. This approximation can be rigorously formulated on an ideal lattice of
infinite Hausdorff dimension, i.e, the Bethe lattice shown in Fig.~\ref{Fig:Tree}(a) \cite{Ostilli2012}. The Bethe lattice is also intimately connected with the dynamical mean-field theory \cite{Eckstein2005,Olga2016,Semerjian2009}.

A Bethe lattice is a loop-free graph where each site is
connected to $z$ neighbours, i.e., $z$ is the coordination number.
Figure~\ref{Fig:Tree}(a), as an example, shows the structure of a $z=3$ Bethe
lattice, whose lattice sites are all equivalent and there exists no boundary on this infinite lattice. This Bethe lattice resembles a two-dimensional honeycomb lattice
locally [as emphasized in Fig.~\ref{Fig:Tree}(b)], 
but it does not contain any closed loops, e.g., hexagons.
A Bethe lattice becomes a Cayley tree if the lattice size is finite,
where the sites are arranged in shells around a root site. 
In stark contrast to the Bethe lattice, there exist boundary sites, which are 
as many as the bulk sites, on the $z=3$ Cayley tree.

Tensor networks provide efficient and accurate representations of quantum
manybody states both at zero and finite temperatures. The simple update \cite%
{Jiang2008}, together with many other optimization schemes \cite%
{Verstraete2008,Orus2014}, has been widely adopted in the tensor-network
simulations of quantum lattice models \cite{Li2012,Li2010,Liu2014,Liu2015,Liu2016}. It has also been shown that the simple update \cite{Jiang2008}, 
or more rigorously the canonical tree tensor
network (TTN) method, is numerically exact on the Bethe lattice \cite{Li2012}. 
In addition, the density matrix renormalization group (DMRG) has also been
employed to study the magnetic orders and other physical properties of the
Heisenberg model on the Cayley tree \cite{Otsuka1996,Friedman1997,Kumar2012,Changlani2013,Changlani2013PRL}. 
Besides, the Bethe approximation has also been used
in investigating the Fermi-Hubbard systems, 
where the single-particle Green's function as well as the density of states  
are calculated \cite{Brinkman1970,Kittler1976,Brouers1982}. 
However, most of these studies are restricted in the $T=0$ properties,
and there have not much investigations on the thermodynamic properties of quantum
lattice models on the Bethe lattice.

In this paper, we extend the TTN approach from zero to finite
temperatures and show that it provides an efficient and accurate method to
simulate the thermodynamic properties
of the Bethe-lattice quantum lattice models. 
Through the calculations of magnetic order parameters, 
we obtain the finite-temperature phase diagram of the
anisotropic Heisenberg model. Similar as in the two-dimensional honeycomb
lattice, three different magnetic ordered phases, i.e., the ferromagnetic
(FM), antiferromagnetic (AF), and planar XY phases, are found on the Bethe lattice.
The planar XY phase is separated from the FM and AF
phases by two first-order phase transition lines at $\Delta =-1$ and $1$,
respectively, again resembling the two-dimensional case \cite{Wind2004}.

The correlation length, as shown in Ref. \cite{Li2012}, is finite on a Bethe
lattice. Here we show that a thermal phase transition can nevertheless happen 
when the correlation length reaches a ``critical" value $\xi_c = 1/\ln{(z-1)}$
on the Bethe lattice. As revealed by the temperature dependence of
thermodynamic quantities in low temperatures, the low-energy excitations of
the XXZ model are always gapped. We propose a linear spin wave theory (LSWT)
on the Bethe lattice, which gives insight into the low-energy
excitations of the system and provides good estimates of the magnon gaps.

The paper is organized as follows. In Sec.~\ref{Sec:MoMe}, we briefly
introduce the Heisenberg XXZ model and the canonical TTN method on the Bethe
lattice. The results obtained with this method are presented in Sec.~\ref{Sec:Bethe}. 
In Sec.~\ref{Sec:LSWT}, we present a spin wave analysis
of the model based on the $q$-representation. Finally, in Sec.~\ref{Sec:Summary}, 
we summarize and discuss about further applications of the
method introduced in this work.

\section{Interacting Spin Model and Tensor Network Method}

\label{Sec:MoMe}

\subsection{The Heisenberg XXZ model}

The Hamiltonian of Heisenberg XXZ model reads
\begin{equation}
H=\sum_{\langle i,j\rangle }(S_{i}^{x}S_{j}^{x}+S_{i}^{y}S_{j}^{y}+\Delta
S_{i}^{z}S_{j}^{z})-h\sum_{i}S_{i}^{z},  \label{Eq:XXZ}
\end{equation}%
where $\Delta $ is the anisotropic coupling constant, and $\langle i,j\rangle $ 
denotes a pair of nearest-neighboring sites. $h$ represents an
external magnetic field, which is set as zero if not mentioned explicitly. 
Note the model with $\Delta =-1$ is equivalent to 
the isotropic FM Heisenberg model upon a $\pi$-rotation
around the $z$-axis on one of the two sublattices, 
labelled by green and orange colors in Fig.~\ref{Fig:Tree}(a), respectively.
In the discussions below, we concentrate on the XXZ model defined on a
Bethe lattice of coordination number $z=3$. 

\subsection{Canonical TTN representation of the density matrix}

As shown in Fig.~\ref{Fig:Tree}(b), the density matrix on the $z=3$ Bethe
lattice can be expressed as a TTN. A local tensor $T$ defined at a node
contains two physical bonds, labelled by $\sigma$ and $\sigma^\prime$,
representing respectively the initial and final states of the density matrix.
There are three geometric bonds, labelled by $\zeta$ ($=x,y,z$), through which the network is connected. Besides this, there is also a vector or diagonal matrices $\lambda_\zeta$ defined on each internal bond $\zeta$, whose square (i.e., $\lambda_\zeta^2$) represents the entanglement
spectra between the two blocks that are separated by this bond.

Given a density matrix, its TTN representation is not uniquely determined.
By inserting a pair of invertible matrices, $U$ and its inverse
$U^{-1}$, on each internal bond, we do not alter the density matrix. Hence, the
TTN representations contain huge gauge redundancy. 
Nevertheless, the gauge degree of freedom can be fixed by converting a
TTN into a canonical form, in which the local tensors $T$'s and diagonal
bond matrices $\lambda$'s satisfy a set of canonical equations. Taking the $%
z $ bond as an example, the canonical equation is
\begin{equation}
\sum_{x,y=1}^{D} \sum_{\sigma,\sigma^\prime=1}^d \lambda_x^2 \lambda_y^2
(T_\alpha^*)_{xyz}^{\sigma\sigma^\prime}
(T_\alpha)_{xyz^\prime}^{\sigma\sigma^\prime} = \mathbb{I}_{zz\prime},
\label{Eq:CanonCond}
\end{equation}
where $\alpha=A$ or $B$, representing the two sublattices of the $z=3$ Bethe
lattice, $d=2$ is the dimension of the local basis states of spin-1/2, and $\mathbb{I}$ is a $D\times D$ identity matrix, with $D$ the dimension of
geometric bond. 
The canonical equations along the $x$ and $y$ bonds can be obtained
from Eq. (\ref{Eq:CanonCond}) through a cyclic permutation of $x$, $y$, and $%
z$. This kind of canonical form has already been used in one-dimensional
tensor network algorithms, including the density matrix renormalization
group (DMRG) \cite{PhysRevLett.69.2863}, time-evolution block decimation
\cite{Vidal2007,OrusVidal2008}, and linearized tensor renormalization group
\cite{Li.w+:2011:LTRG, Dong.y+:2017:BiLTRG}, etc.

Given an arbitrary TTN representation which generically does not satisfy these canonical equations, we can nevertheless
gauge the TTN into the canonical form through a so-called canonicalization
procedure elaborated in App.~\ref{App:CanonBethe}.

\subsection{Imaginary-time evolution}

\label{SSec:ITE_BL}

The tensor network representation of thermal density matrix $\rho (\beta )$ 
\cite{Li.w+:2011:LTRG,Ran.s+:2012:Super-orthogonalization,Chen.b+:2017:SETTN,
Dong.y+:2017:BiLTRG,Chen2018,Czarnik.p+:2012:PEPS,Czarnik.p+:2015:PEPS,Czarnik.p+:2016:TNR,Czarnik.p+:2017:Sign,Orus2018,Czarnik2019a,Czarnik2019b}
can be determined by taking an imaginary-time evolution, 
starting from an infinitely high temperature at
which $\rho (\beta )$ is represented by the identity operator. 
This is achieved by taking a Trotter-Suzuki decomposition for the density matrix
\begin{eqnarray}
\label{Eq:TS_BL}
\rho (\beta ) &=&e^{-\beta H}\approx (e^{-\tau H_{x}}e^{-\tau H_{y}}e^{-\tau
H_{z}})^{N}, \\
H_{\zeta } &=&\sum_{i}h_{i,i+\zeta },\text{ \ \ \ \ \ \ }\left( \zeta
=x,y,z\right) ,
\end{eqnarray}
where $\beta \equiv 1/T$ is the inverse temperature and $\tau =\beta /N$. 
$h_{i,i+\eta }$ is the interacting Hamiltonian between (nearest-neighboring) 
sites $i$ and $i+\zeta$ along $\zeta$(=$x,y,z$) directions. 
In practical calculations, the Trotter step $\tau $ is set as a small
value, e.g., $\tau =0.01$, to control the Trotter error. For the model we
study, all the local terms in each $H_{\eta }$ commute with each other,
i.e., $[h_{i,i+\zeta },h_{i^{\prime },i^{\prime }+\zeta }]=0$, thus we can
further decompose $\exp \left( -\tau H_{\zeta }\right) $ into a product of
local evolution gates, i.e.,
\begin{equation*}
P_{\zeta }=\exp \left( -\tau H_{\zeta }\right) =\prod_{i}\exp \left( -\tau
h_{i,i+\zeta }\right) .
\end{equation*}
A detailed introduction to the update scheme of local tensors in the
imaginary-time evolution is given in App.~\ref{App:SimpleU}. 

  We dub the simple update equipped with the canonicalization procedure (App.~\ref{App:CanonBethe}) as \textit{canonical} TTN scheme on the Bethe lattice. The canonical TTN approach adopted in the present work indeed improves the accuracy and stability of the calculations, especially for the case with a thermal phase transition, as benchmarked in App. \ref{App:SvsC}.

\begin{figure}[!tbp]
\includegraphics[angle=0,width=0.85\linewidth]{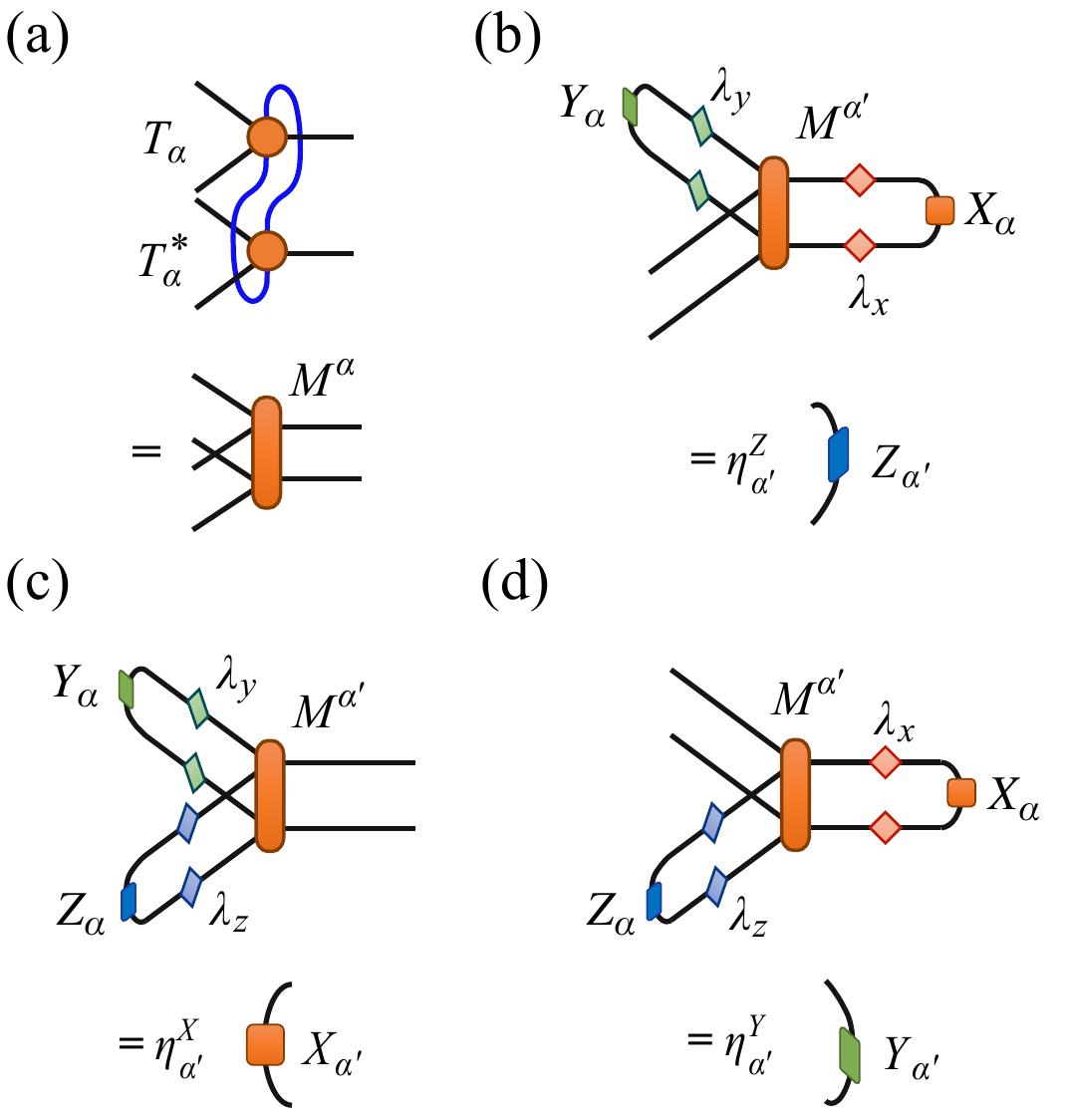}
\caption{(Color online) The generalized eigenvalue problems of transfer
tensor $M^\protect\alpha$. The construction of $M^\protect\alpha$ is
illustrated in (a), whose dominant eigenvalues can be obtained through
iterative tensor contractions following (b-d), see the main text for
details. }
\label{Fig:Contraction}
\end{figure}

  After obtaining the tensor network representation, to evaluate thermodynamic quantities, we notice that the density matrix opertor $\rho (\beta )$ is a product of two half density matrix operators $\rho (\beta /2)$
\begin{equation}
\rho (\beta )=\rho \left( \beta /2\right) \rho \left( \beta /2\right) =\rho
^{\dagger }\left( \beta /2\right) \rho \left( \beta /2\right) .
\end{equation}%
This TTN operator $\rho (\beta /2)$ can be also regarded as a
\textquotedblleft supervector" $|\rho (\beta /2)\rangle $, defined in the
(enlarged) product space of the \textquotedblleft ket" and \textquotedblleft
bra" spaces. In this supervector representation, the partition function then
becomes an inner product of $|\rho (\beta /2)\rangle $ and its vector
conjugate, namely
\begin{equation}
Z(\beta )=\mathrm{Tr}\rho (\beta )=\langle \rho (\beta /2)|\rho (\beta
/2)\rangle .
\end{equation}
Thus the partition function can be obtained by contracting a bilayer TTN
\cite{Dong.y+:2017:BiLTRG}.

On the Bethe lattice, contracting the above bilayer supervectors is
equivalent to solving the dominant eigen-problem of the transfer tensor $%
M^{\alpha }$ defined by
\begin{equation}
M_{xx^{\prime };yy^{\prime };zz^{\prime }}^{\alpha }=\sum_{\substack{ \sigma
_{1},\sigma _{2}}}(T_{\alpha })_{xyz}^{\sigma _{1}\sigma _{2}}(T_{\alpha
}^{\ast })_{x^{\prime }y^{\prime }z^{\prime }}^{\sigma _{1}\sigma _{2}},
\label{Eq:TransT}
\end{equation}%
whose schematic representation is shown Fig.~\ref{Fig:Contraction}(a). Supposing that $X_{\alpha }$, $Y_{\alpha }$ and $Z_{\alpha
} $ are the three dominant eigenvectors to find, $M^{\alpha }$ ($\alpha =A$
or $B$) simply transfers two of these eigenvectors in one of the
sublattice to an eigenvector in the other sublattice, see Figs.~\ref%
{Fig:Contraction}(b-d). For example, $M^{B}$ transfers $X_{A},Y_{A}$ to $%
Z_{B}$ through the equation [Fig.~\ref{Fig:Contraction}(b)],
\begin{equation}
\sum_{\substack{ xx^{\prime }yy^{\prime }}}(X_{A})_{xx^{\prime
}}(Y_{A})_{yy^{\prime }}\Lambda_{xx^{\prime }yy^{\prime
}}M_{xx^{\prime },yy^{\prime },zz^{\prime }}^{B}=\eta
_{B}^{Z}(Z_{B})_{zz^{\prime }},
\label{Eq:Iter_BL}
\end{equation}
where $\Lambda_{xx^{\prime }yy^{\prime }}=\lambda _{x}\lambda
_{x^{\prime }}\lambda _{y}\lambda _{y^{\prime }}$. This defines a
generalized eigenvalue problem associated with the transfer tensor 
$M^{\alpha }$ and can be solved iteratively, 
i.e., we start from six random vectors and 
perform the contractions until all vectors converge.

In stark contrast to the conventional linear eigenvalue problem, there is a gauge
flexibility in the definition of eigenvalues $\eta_\alpha$ here. 
Once the normalization condition of eigenvectors $X_{\alpha }$, 
$Y_{\alpha }$ and $Z_{\alpha} $ is varied, 
the corresponding eigenvalues $\eta _{\alpha }^{X,Y,Z}$ also change, 
i.e, they are not uniquely defined.
In the following, we fix the gauge by normalizing 
all the dominating eigenvectors to 1, e.g., 
$\lVert X_{\alpha }\rVert =\sqrt{\mathrm{Tr}(X_{\alpha }^{\dagger}X_{\alpha })}=1$. 
This is a convenient choice, 
and note that once the density matrix $\rho \left( \beta /2\right)$ 
is in the canonical form, c.f., Eq.~(\ref{Eq:CanonCond}), 
the above iterative contraction procedures can be skipped,
since each dominant eigenvector is just identity matrices 
if represented as a $D\times D$ matrix.

Given the eigenvectors $X_{\alpha }$, $Y_{\alpha }$ and $Z_{\alpha}$, 
we can evaluate the thermal expectation values of local operators, such as the
local magnetization. For example, to evalueate the expectation value of an
operator $\hat{O}$ on the $A$-sublattice, we first construct the single-site
reduced density matrix
\begin{eqnarray}
(\rho _{A})_{\sigma _{1}\sigma _{2}} &=&\sum_{xx^{\prime }yy^{\prime
}zz^{\prime }\sigma }(X_{B})_{xx^{\prime }}(Y_{B})_{yy^{\prime
}}(Z_{B})_{zz^{\prime }}  \notag \\
&&(T_{A})_{xyz}^{\sigma _{1}\sigma }(T_{A}^{\ast })_{x^{\prime }y^{\prime
}z^{\prime }}^{\sigma _{2}\sigma }\lambda _{x}\lambda _{x^{\prime }}\lambda
_{y}\lambda _{y^{\prime }}\lambda _{z}\lambda _{z^{\prime }}.
\label{Eq:RDM_BL}
\end{eqnarray}
The expectation value is then given by
\begin{equation}
\langle \hat{O}\rangle _{\beta }=\frac{\mathrm{Tr}(\rho _{A}\hat{O})}{%
\mathrm{Tr}\rho _{A}},
\end{equation}
which no longer depends on how the generalized dominant eigenvectors 
are normalized, since $\rho _{\alpha }$ appears in both the
numerator and the denominator. Similarly, we can evaluate the bond energies
from the two-site reduced density matrix, etc.

From the canonical TTN, we can also calculate the bipartite-entanglement
entropy $S_{E}$ using the normalized entanglement spectrum $\lambda _{\zeta }$, as
\begin{equation}
S_{E}=-\mathrm{Tr}\left( \lambda _{\zeta }^{2}\ln \lambda _{\zeta }^{2}\right),
\label{Eq:Se}
\end{equation}
which reflects both the quantum entanglement and classical correlations in
a thermal equilibrium state. For gapless systems, the entanglement $S_E$
might exhibit a universal logarithmic scaling as a function of temperatures at low $T$, 
e.g., in the one-dimensional quantum critical points and
two-dimensional Heisenberg models \cite
{Prosen.t+:2007:Entropy,Barthel.t:2017:FiniteT,Dubail17,Chen2018,Chen2018b}.

Therefore, it is of great interest to explore the scaling behaviors of $S_{E}$, particularly near the phase transition temperatures, 
for the XXZ model on the Bethe lattice. Note, $S_{E}$ provides 
a quantitive measure of the bond dimension that is needed for an
accurate representation of the thermal density matrix, especially in low
temperatures. Generically, the bond dimension $D$ scales exponentially with 
$S_{E}$, which would be saturated at low temperatures in a gapped system.
On the contrary, in a gapless system $D$ would scale algebraically 
with inverse temperature $\beta$, given that $S_E \sim \ln \beta$.

\section{Numerical results}

\label{Sec:Bethe}

Here we present the thermodynamic results calculated with the canonical TTN
method. In our calculations, up to $D=80$ states are retained in the
geometric bonds of local tensors to ensure that the results are converged down
to $T=0.05$.

To benchmark the method, we have evaluated 
the thermodynamic quantities with the
canonical TTN in the classical limit $\Delta =\infty $ [see Eq.~(\ref{Eq:XXZ}%
)], at which the XXZ model is reduced to the exactly soluble Ising model. In
this limit, the density matrix has an exact TTN representation with
a bond dimension $D=2$. We find that our numerical results agree excellently
with the exact values. One can refer to App.~\ref{App:Ising} for details. Below, we show our TTN results of the quantum XXZ model at finite $T$.

\subsection{Phase diagram}

\label{SSec:SymBreak} It has been well-established that on a two-dimensional
bipartite lattice, say, the honeycomb or square lattice, the XXZ model with $%
-1\leq \Delta \leq 1$ breaks the continuous symmetry and possesses a
long-range order in the ground state. However, at finite temperatures, the
continuous symmetry is restored and the long-range magnetic order is melted
by the low-lying excitations according to the Mermin-Wagner theorem \cite%
{Mermin-Wagner}.

\begin{figure}[!tbp]
\includegraphics[width=0.9\linewidth]{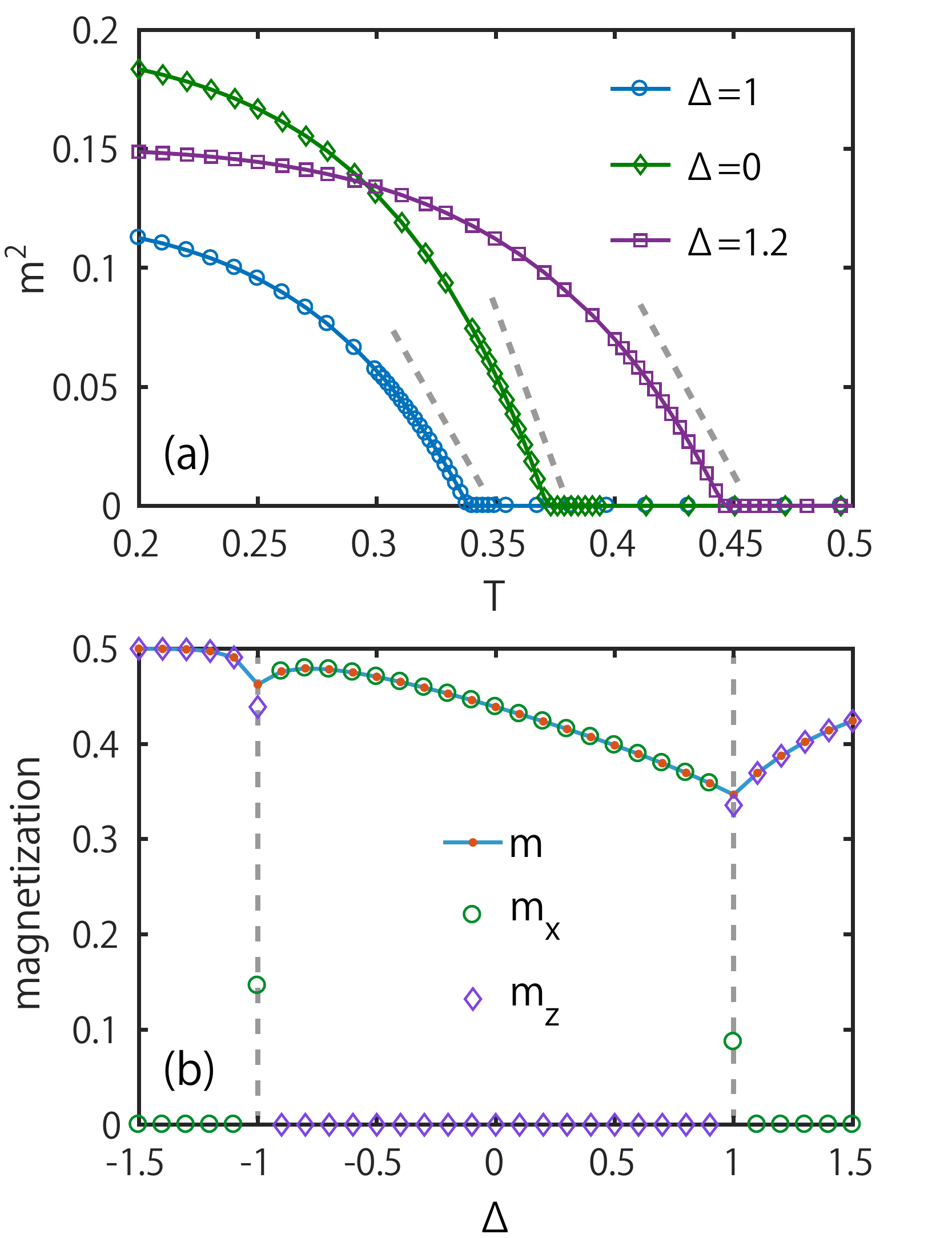}
\caption{(Color online) (a) The spontaneous magnetization of the
Bethe-lattice XXZ model, where $m^2 \neq 0$ for $T<T_c$ and vanishes for $T
\geq T_c$. The dashed lines emphasize a linear relation of $m^2$ vs. $T$
near $T_c$. (b) The magnetization $m$ and its 
two components $m_x$ and $m_z$ are plotted vs. $\Delta$, 
at a low temperature $T=0.125$. 
The two vertical dashed lines denote 
the high symmetry points $\Delta=\pm 1$.}
\label{Fig:MagBL}
\end{figure}

\begin{figure}[!tbp]
\includegraphics[width=0.9\linewidth]{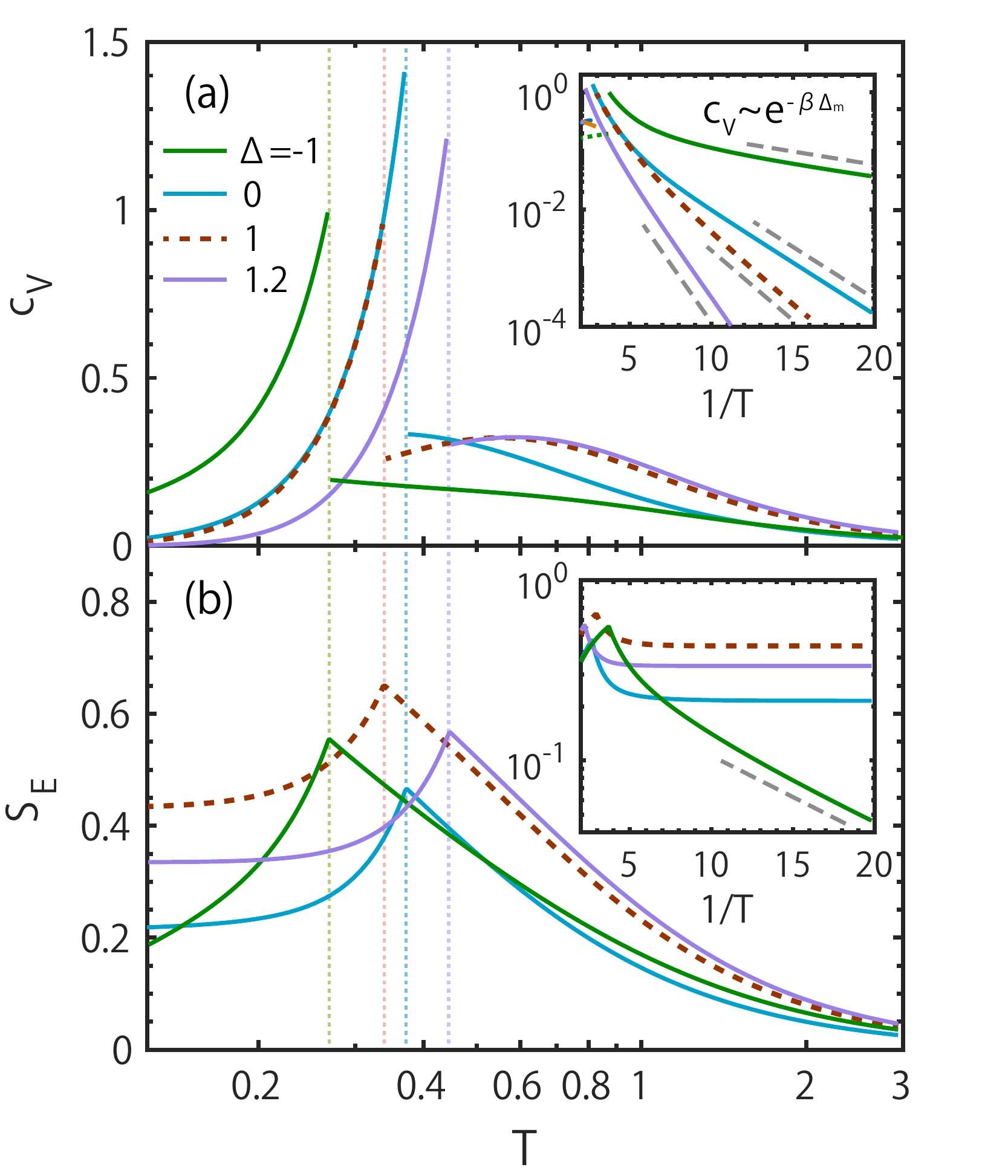}
\caption{(Color online) Temperature dependence of (a) the specific heat $c_V$
and (b) the entanglement entropy $S_E$ for the Heisenberg XXZ model. $c_V$
drops exponentially with $T$ in low temperatures. This exponential decay is emphasized by the dashed gray lines in the inset of (a) where the low temperature part is zoomed in. $S_E$ shows a cusp at $T_c$ and converge to a finite value at low
$T$ [the inset of (b)]. In the FM state, $\Delta\le -1$, $S_E$ drops
exponentially to zero in the zero temperature limit, as indicated by the dashed gray line in the inset of (b). The vertical dashed lines in both (a) and (b) represent the critical temperatures of various $\Delta$, with the 
specific $T_c$ values shown below in Fig.~\ref{Fig:CorrLen}.}
\label{Fig:MeanFieldPT}
\end{figure}

Figure~\ref{Fig:MagBL} shows the temperature dependence of the magnetic
order parameter
\begin{equation}
m=\sqrt{m_{x}^{2}+m_{z}^{2}},
\end{equation}%
where $m_{x}=\langle S_{i}^{x}\rangle _{\beta }$ and $m_{z}=\langle
S_{i}^{z}\rangle _{\beta }$ are the components within and perpendicular to
the XY plane ($\langle S_{i}^{y}\rangle _{\beta }=0$ by default), respectively. 
The absolute values of $m$ are the same on the $A$ and $B$ 
sublattices, and there is no
spontaneous symmetry breaking, i.e., $m=0$, in high temperatures. 
However, $m$ becomes finite when the temperature drops below a critical value $T_{c}$.  Particularly, we find that $m^{2}$ varies linearly 
with temperature just below $T_{c}$, i.e.,
\begin{equation*}
m^{2}\propto \frac{T_{c}-T}{T_{c}}.
\end{equation*}
We have checked three cases shown in Fig.~\ref{Fig:MagBL}(a) with $\Delta =0$, $1$ and $1.2$, which all fall into this scalings in the vicinity of $T_c$, 
thus indicating that the transition is mean-field-like.

Figure \ref{Fig:MeanFieldPT}(a) shows the temperature dependence of the
specific heat at several different $\Delta$ values. A $\lambda$-jump is observed
at the critical point $T_{c}$, which confirms that the transition from the
paramagnetic to magnetic-ordered phase is continuous 
(in a mean-field-like fashion). 
On the other hand, as shown in 
Fig.~\ref{Fig:MeanFieldPT}(b), the entanglement 
entropy curve vs. temperature exhibits a cusp, 
rather than a diverging peak, at $T_c$. The absence of
divergent $S_E$ at $T_c$ is a natural consequence of the finite correlation
length $\xi$ in the system (see discussions on $\xi$ 
in Sec.~\ref{SSec:Magnon} below). 
It allows us to perform accurate thermal simulations down to low
temperatures, $T\leq T_c$, by retaining a finite number $D$ of bond states.

\begin{figure}[!tbp]
\includegraphics[width=0.9\linewidth]{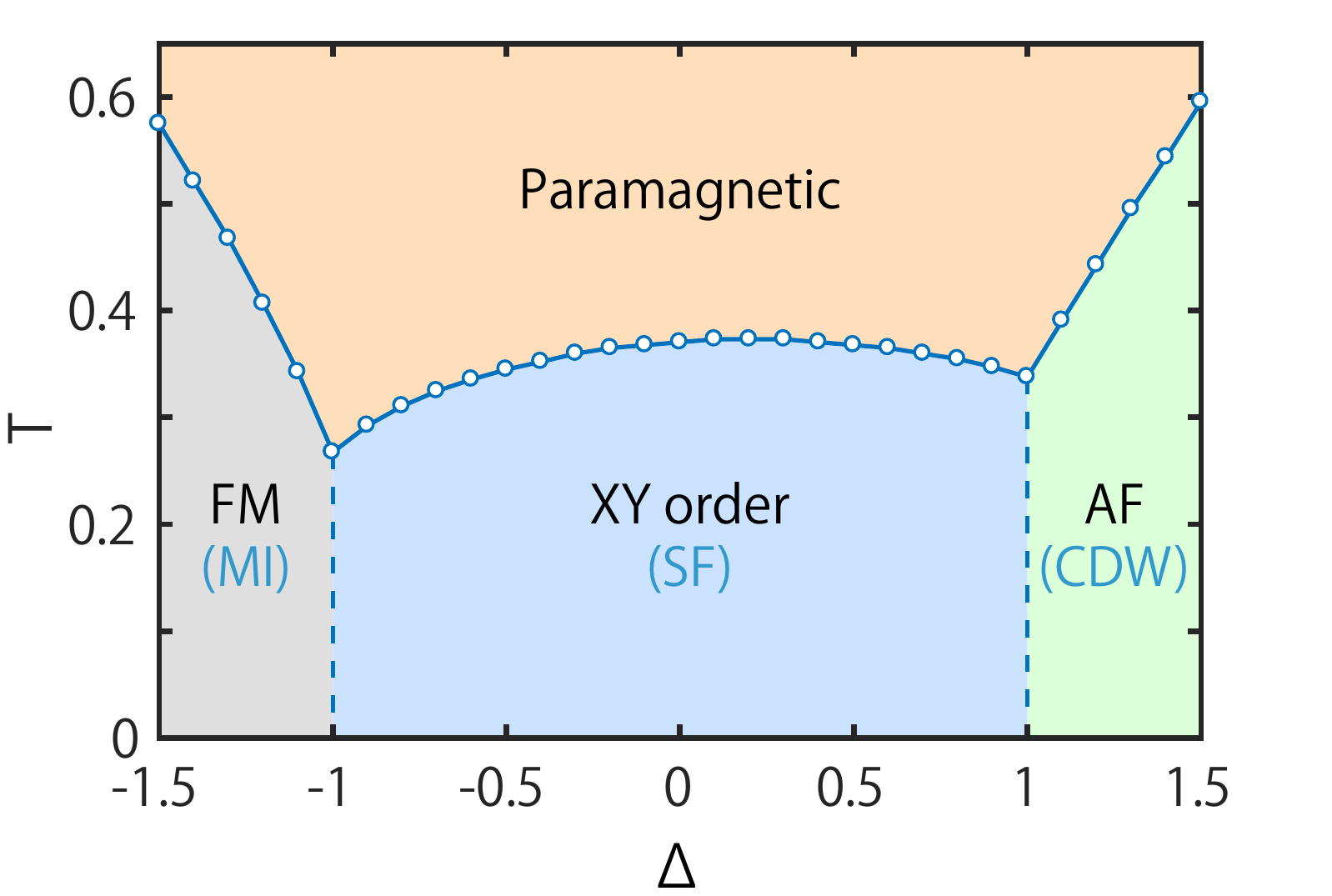}
\caption{(Color online) $\Delta$-$T$ phase diagram of the Bethe-lattice XXZ
model, which contains the FM, XY and AF ordered phases in low temperatures.
The solid blue lines (with circle symbols) represent the second-order phase
boundaries between the ordered and the paramagnetic phases. In the brackets
are corresponding phases in the hardcore boson language, including the SF,
CDW and MI phases (see main texts).}
\label{Fig:PhaseDiagram}
\end{figure}

Figure~\ref{Fig:PhaseDiagram} shows the $T$-$\Delta $ phase diagram of the Heisenberg
XXZ model on the Bethe lattice, where three magnetic ordered
phases are observed in low temperatures, by varying 
the anisotropic parameter $\Delta $. 
As depicted in Fig.~\ref{Fig:MagBL}(b), for $\Delta >1$ and $\Delta <-1$,
the low-temperature states are AF and FM ordered, respectively.
The system spontaneously breaks the $Z_{2}$ symmetry, 
as characterized by $m_{z}\neq 0$ and $m_{x}=0$. 
When $|\Delta |<1$, the planar U(1) symmetry is
broken, with $m_{x}\neq 0$ and $m_{z}=0$ in low temperatures. 
This is in stark contrast to the corresponding 
two-dimensional lattice models with the
same $\Delta $ parameter, 
where no long-range order exists at any finite
temperature according to the Mermin-Wagner theorem. 

On the two vertical phase boundaries 
(marked by the two dashed lines in Fig.~\ref{Fig:PhaseDiagram}), 
both $m_{x}$ and $m_{z}$ become finite, but their ratio is
somewhat arbitrary. This indicates that the system is in a random mixture of $%
Z_{2}$-symmetry-breaking AF (or FM) and U(1)-symmetry-breaking XY phases. In
other words, the U(1)$\otimes Z_{2}$ symmetry at $\Delta =-1$ or the SU(2)
symmetry at $\Delta =1$ is broken, and the transitions from the planar-XY
order to either FM- or AF-ordered phase is of the first order. Both $m_{x}$
and $m_{z}$ become discontinuous at $\Delta =\pm 1$, while
$m$ remains continuous across these two high-symmetry points. 

\begin{figure}[!tbp]
\includegraphics[width=0.88\linewidth]{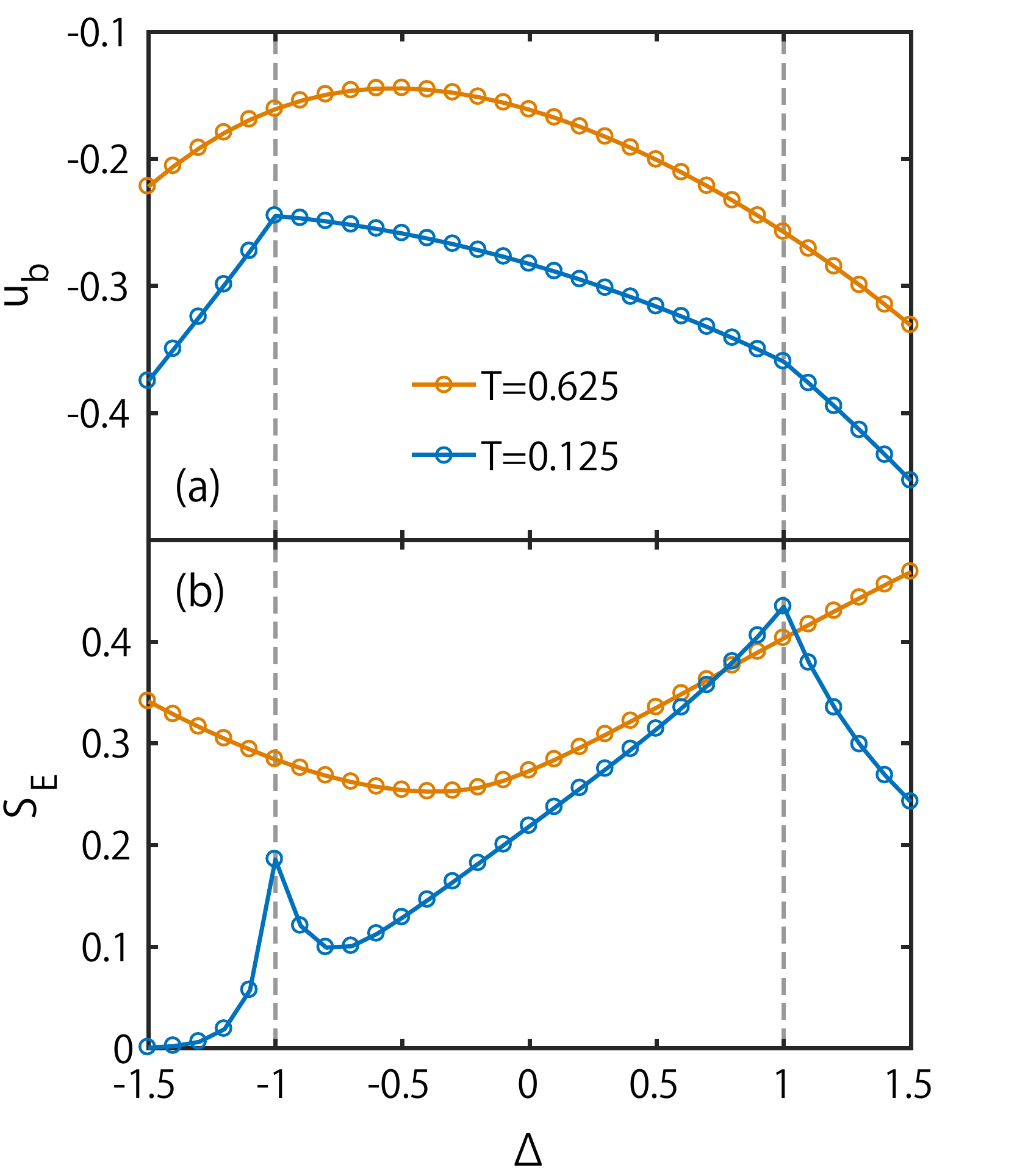}
\caption{(Color online) (a) Internal energy per bond $u_b$ and (b)
entanglement entropy $S_E$ vs. $\Delta$ 
at two temperatures, $T=0.125$ and $0.625$.
The two vertical dashed lines denote the high symmetry points $\Delta=\pm 1$.}
\label{Fig:PhaseSP}
\end{figure}

These two first-order phase transition lines can also be identified via
other thermal measurements.
Figure~\ref{Fig:PhaseSP}(a) shows the energy per bond $u_{b}$ as a function
of $\Delta $ in the high-temperature paramagnetic phase with $T=0.625$, and deep in the symmetry-breaking phase with $T=0.125$. At $T=0.125$, the $%
u_{b}$ curve shows two turning points at $\Delta =\pm 1$, where the
derivative $\partial u_{b}/\partial \Delta $ becomes discontinuous, as a
consequence of the first-order phase transitions. On the other hand, at $T=0.625$, both $u_{b}$ and $\partial u_{b}/\partial \Delta $ vary smoothly, suggesting the absence of phase transitions vs. $\Delta$.

Similarly, the phase transitions among these three phases can be seen from
the $\Delta $-dependence of the entanglement entropy $S_{E}$. As shown in
Fig.~\ref{Fig:PhaseSP}(b), $S_{E}$ exhibits two peaks at $\Delta =\pm 1$
at low temperature $T=0.125$, which also becomes smooth at a high value $T=0.625$.

Lastly, the XXZ spin model can be mapped onto a hardcore boson model with nearest-neighboring (NN) interactions, by setting $S_{i}^{-}=b_{i}$ and $%
S_{i}^{z}=b_{i}^{\dagger }b_{i}-1/2$, where $b_{i}$ is a hardcore boson
operator. In the boson language, the XY phase corresponds to a superfluid
phase (SF) with an off-diagonal long-range order, and the FM and AF phases
correspond to a Mott insulator (MI) and a charge density wave (CDW) phases,
respectively. Along the two vertical phase boundary lines from low
temperatures up to $T_{c}$, the SF phase coexists with the MI or CDW phases.

\subsection{Low-lying excitations and quasi-Goldstone modes}
\label{SSec:Magnon}

From the inset of Fig.~\ref{Fig:MeanFieldPT}(a), it is clear that the
specific heat $c_{V}$ decays exponentially with $\beta $ in low temperatures,
no matter in which magnetically ordered phase, i.e.,
\begin{equation}
c_{V}\sim e^{-\beta \Delta _{m}}.
\end{equation}%
This indicates that there is a finite excitation gap in the low-lying energy
spectrum, quantified by the exponent $\Delta _{m}$ in the above equation.
The values of $\Delta _{m}$ (Table~\ref{Tab:MGap}) can be obtained by
fitting the low-$T$ results of the specific heat $c_{V}$ or the internal
energy $u$. For example, for the FM phase at $\Delta =-1^{-}$, the
low-temperature internal energy is approximately described by the formula
\begin{equation}
u(\beta )-u_{0}\simeq g(\beta )e^{-\beta \Delta _{m}}
\end{equation}%
where $u_{0}=-0.375-h/2$ is the ground state energy per site, 
and $h$ is an external magnetic field, and $g(\beta)$
is some polynomial prefactor. From the derivative of this equation
\begin{equation*}
\frac{d\mathrm{ln}[u(\beta )-u_{0}]}{d\beta }=\frac{1}{g}\frac{dg}{d\beta }%
-\Delta _{m},
\end{equation*}%
and through a polynomial fitting, 
the magnon gap $\Delta _{m}$ can be
readily read out from the intercept at $T=1/\beta=0$. 
Moreover, for this FM system, one can
further tune the magnon gap by changing the external magnetic
fields $h$. The FM excitation gaps of
various (small) magnetic fields $h=0,0.1,0.2$ are also obtained by fitting
the internal energy curves and shown in Table~\ref{Tab:MGap}, 
from which we find that $\Delta _{m}(h)-\Delta _{m}(0)\simeq h$.
This suggests that the elementary excitations in the FM phase
are magnons with spin $S=1$.

To further clarify the nature of low-lying excitations, we have evaluated
the correlation length from the transfer matrix along the path 
$... \, B \overset{z}{\rightarrow }A\overset{x}{\rightarrow }B
\overset{z}{\rightarrow }...$ on the Bethe lattice, meaning going
through the $z$ bond of the $T_B$ tensor to $T_A$, 
and then through its $x$ bond to the next $T_B$, 
and so on. There are also other paths that can be used to define
the transfer matrix. But all these paths are physically equivalent and the
correlation lengths obtained thereof are also found exactly the same.

Suppose $\eta _{0}$ and $\eta _{i}$ are the dominant and the ($i+1$)'th largest
eigenvalue of the transfer matrix, the correlation length can be determined as 
\begin{equation}
\xi _{i}=2\ln ^{-1}\left\vert \frac{\eta _{0}}{\eta _{i}}\right\vert,
\label{Eq:CorrLen}
\end{equation}%
where $\xi _{1}$ is the largest correlation
length the system can have, and $\xi _{i>1}$ is related to a shorter-ranged
correlation function. 

\begin{table}[!tbp]
\caption{Magnon gaps of the XXZ model with different anisotropies $\Delta$
and external fields $h$. The numerical and LSWT results are compared, where
the numbers in the parentheses in the TTN data are numerical fitting errors.}
\label{Tab:MGap}
\centering
\begin{tabular}{cccc}
\hline\hline
Model \,\,\, & ($\Delta$, $h$) \,\,\, & $\Delta_m$ (TTN, fitted) \,\,\, & $\Delta_m$ (LSWT, $z=3$) \\ \hline
AF \,\,\, & (1, 0) \,\,\, & 0.56(3) & $0.5$ \\ \hline
XY \,\,\, & (0, 0) \,\,\, & 0.41(1) & $0.3587$ \\ \hline
FM \,\,\, & (-1, 0) \,\,\, & 0.085(3) & $0.0858$ \\
& (-1, 0.1) \,\,\, & 0.186(1) & $0.1858$ \\
& (-1, 0.2) \,\,\, & 0.286(2) & $0.2858$ \\ \hline\hline
\end{tabular}%
\end{table}

\begin{figure*}[!tbp]
\includegraphics[width=1\linewidth]{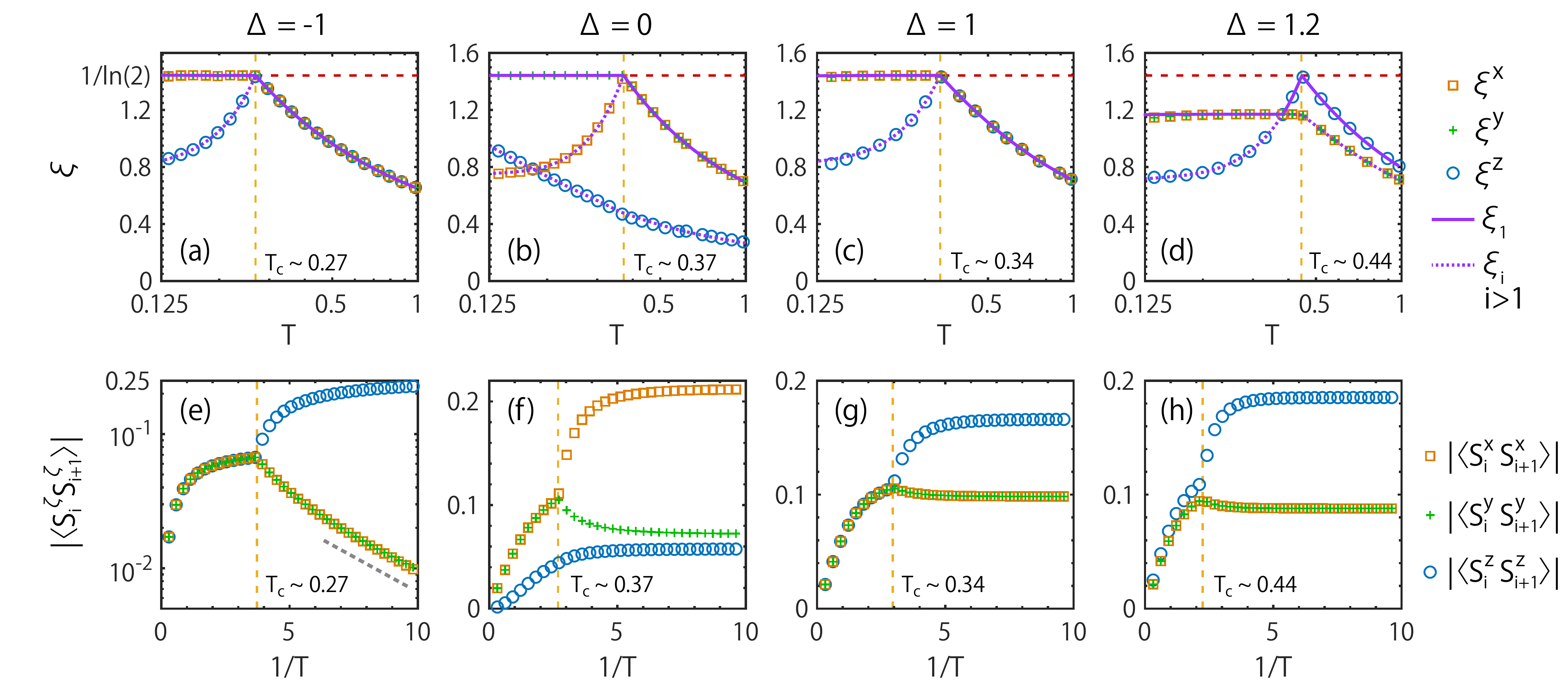}
\caption{(Color online) (a-d) Correlation length $\protect\xi$ vs. $T$ and  (e-h) the absolute value of corresponding NN correlations 
$|\langle S_i^\zeta S_{i+1}^\zeta \rangle |$ ($\zeta=x,y,z$) vs. $1/T$,
 for $\Delta=-1$, $0$, $1$ and $1.2$, respectively. In (a-d), 
the purple lines \{$\protect\xi_i$\} are extracted from the transfer-matrix spectra, 
in which the solid lines are the leading correlation length $\protect\xi_1$ and the dashed ones are certain shorter correlation length $\protect\xi_i$ [see Eq.~\protect(\ref{Eq:CorrLen}) for definition, $i>1$]. And the markers label the correlation lengths $\protect \xi^{x,y,z}$ fitted from the spin-spin correlators $C^{x,y,z}(l)$. The dashed gray line in (e) illustrates the exponential decay vs. $1/T$, as a guide to the eye. For $\Delta=\pm1$ in (a), (c), (e) and (g), we have chosen the $z$ axis as where the spontaneous magnetization takes place and $\protect\xi^{x,y}$ are related to transverse fluctuating modes on the XY plane. 
The vertical dashed (yellow) lines in all panels represent the critical temperatures $T_c$ determined in Fig.~\protect\ref{Fig:MeanFieldPT}, and the horizontal dashed
(red) lines in (a-d) denote the maximal 
correlation length $\protect\xi_c=1/\ln(2) $. }
\label{Fig:CorrLen}
\end{figure*}

Alternatively, one can also estimate the correlation length $\xi$ directly
from the real-space spin-spin correlation function
\begin{equation}
C^{\zeta }(l)=\left\langle S_{i}^{\zeta }S_{i+l}^{\zeta }\right\rangle
-\left\langle S_{i}^{\zeta }\right\rangle \left\langle S_{i+l}^{\zeta
}\right\rangle ,\text{ \ \ \ \ \ \ (}\zeta =x,y,z\text{)}
\end{equation}%
where $l$ measures the distance between the two sites along the path. 
Given the $C^{\zeta }(l)$ data, the correlation length $\xi$ 
can be obtained by fitting the large-distance correlation
function with an exponentially varying function.

Fig.~\ref{Fig:CorrLen}(a-d) shows the correlation lengths obtained with the
above two approaches. The correlation length $\xi ^{z(x)}$ determined from
the dominating correlators, $C^{z}(l)$ for $\Delta =\pm 1,1.2$ and $C^{x}(l)$
for $\Delta =0$, exhibits cusps exactly at the critical temperatures $T_{c}$. 
The value of the correlation length at the critical point, $\xi _{c}=1/\ln
(2)\simeq 1.4427$, equals the critical upper bound of the
correlation length on the $z=3$ Bethe lattice. As discussed in 
Ref.~\onlinecite{Li2012}, the number 
of spins that correlate with a root spin 
at a given site grows exponentially with their distances.
Therefore, a magnetic order-disorder phase transition with a diverging
magnetic susceptibility happens when $\xi$, 
though being finite, reaches the critical value $1/\ln{(z-1)}$.

For the cases $\Delta =\pm 1$ shown in Fig.~\ref{Fig:CorrLen}(a,c), 
the correlation lengths, $\xi ^{x,y,z}$ along all three directions,
are all equal to the dominant correlation length 
$\xi _{1}$ in the paramagnetic phase above 
$T_{c}$. Below $T_{c}$, the transverse excitation modes remain 
critical and do not change with temperature, i.e., the 
corresponding correlation lengths 
still equal the dominant one, $\xi ^{x,y}=\xi _{1}=\xi_{c}$. 
Moreover, the longitudinal $\xi ^{z}$ drops
immediately below $T_{c}$ with decreasing
temperature, due to the formation of magnetic order along the $z$-direction.
Similar critical behavior has been observed 
in Fig.~\ref{Fig:CorrLen}(b) with $\Delta =0$,
where the degeneracy between $\xi ^{x}$ and $\xi ^{y}$
is broken, since the ordering of spins happens along the $x$-direction. 

For the case $|\Delta |>1$ in Fig.~\ref{Fig:CorrLen}(d), $\xi ^{z}=
\xi _{1}$ is still the dominant correlation length above $T_{c}$,
which reaches the critical value $\xi _{c}$ and drops below $T_{c}$. 
On the other hand, $\xi ^{x,y}$ never reaches the
critical value $1/\ln{2}$, although they do not decay below $T_{c}$,
and surpasses $\xi ^{z}$ at a temperature below $T_{c}$.

The peculiar behaviors of the transverse correlation lengths 
observed in the ordered phases are quite remarkable. 
It indicates that although the true Goldstone modes are absent in the
continuous symmetry breaking phases \cite{Laumann2009}, the transverse
excitations remain \textquotedblleft critical" with a maximal correlation
length $\xi _{c}=1/\ln {(2)}$ on the Bethe lattice. These quasi-critical
transverse excitation modes are reminiscent of the Goldstone modes in a
truly gapless continuous-symmetry-breaking system, 
and they can be regarded as somewhat ``renormalized" Goldstone modes. 

Distinct from the true gapless modes, 
and as the finite correlation lengths imply,
these quasi-Goldstone modes are always gapped, 
and they become activated only 
above certain finite energy scales/temperatures, 
giving rise to the finite-$T$ phase transitions. 
Note that such kind of quasi-Goldstone modes are absent
in the $Z_{2}$-symmetry-breaking phase,
again similar to the two-dimensional lattices where
Goldstone modes are absent for $\left\vert \Delta \right\vert >1$.

As a complementary to correlation length data, 
we show in Figs.~\ref{Fig:CorrLen}(e-h) the absolute value of NN correlators 
$| \langle S_i^\zeta S_{i+1}^\zeta \rangle |$ ($\zeta=x,y,z$). 
In Fig.~\ref{Fig:CorrLen}(e,g), we see distinct behaviors of NN correlators
between the FM ($\Delta =-1$) and AF ($\Delta =1$) cases. 
In the AF phase, $| \langle S_i^{x(y)} S_{i+1}^{x(y)} \rangle |$
approaches a finite value in the zero temperature limit. 
However, in the FM case, $| \langle S_i^{x(y)} S_{i+1}^{x(y)} \rangle |$
decays exponentially as $T$ decreases
and it approaches zero in the $T=0$ limit. This is
consistent with the fact that the FM ground state is simply a direct-product
state $|\uparrow ,\uparrow ,...,\uparrow \rangle $, while the AF ground
state bears quantum entanglement. 
Moreover, as shown in Figs.~\ref{Fig:CorrLen}(f,h),
$| \langle S_i^{x} S_{i+1}^{x} \rangle |$ and $| \langle S_i^{z} S_{i+1}^{z} \rangle |$ are strongest NN correlators in the $\Delta=0$ and $1.2$ cases, 
respectively. In addition, other spin correlators in these two cases 
converge to finite values in the $T=0$ limit.

\section{Linear Spin Wave Theory}

\label{Sec:LSWT}

Here we provide a linear spin wave analysis of the XXZ model on the Bethe
lattice. The results reveal more information on the low-lying excitations,
especially on the quasi-Goldstone mode in the continuous symmetry breaking
phase.

\subsection{The ferromagnetic Heisenberg model}

We first consider the excitations in the FM phase with $\Delta <-1$. By
taking a U(1) rotation $\exp \left( \mathrm{i}\pi S_{n}^{z}\right) $ for all the
spins on the $B$ sublattice, we can transform Eq. (\ref{Eq:XXZ}) into the
form

\begin{equation}
H=\sum_{\langle i,j\rangle }(-S_{i}^{x}S_{j}^{x}-S_{i}^{y}S_{j}^{y}+\Delta
S_{i}^{z}S_{j}^{z})-h\sum_{i}S_{i}^{z}.
\end{equation}%
To carry out the linear spin-wave expansion, we choose an arbitrary site as
a \textquotedblleft root\textquotedblright , labelled as $R_{0}=0$, and then
there exists an unique path that connects it to any other lattice site,  
rendering a convenient way to label the Bethe lattice sites. More
specifically, we label the lattice according to the rule shown in Fig.~\ref%
{Fig:Label_BL}: a site in the $l$-th layer from the root is represented by $%
l $ indices $R_{l}=\left( r_{1}r_{2}...r_{l}\right) $, where $r_{\lambda
}=0,1,...,\tilde{z}-1$ ($\lambda =1,2,..,l$) denotes the way a path can
choose at the $(\lambda -1)$-th branching point. On a $z$-coordinated Bethe
lattice, $\tilde{z}=z$ for the $l=1$ layer and $\tilde{z}=z-1$ for the rest layers.

\begin{figure}[!tbp]
\includegraphics[width=0.95\linewidth]{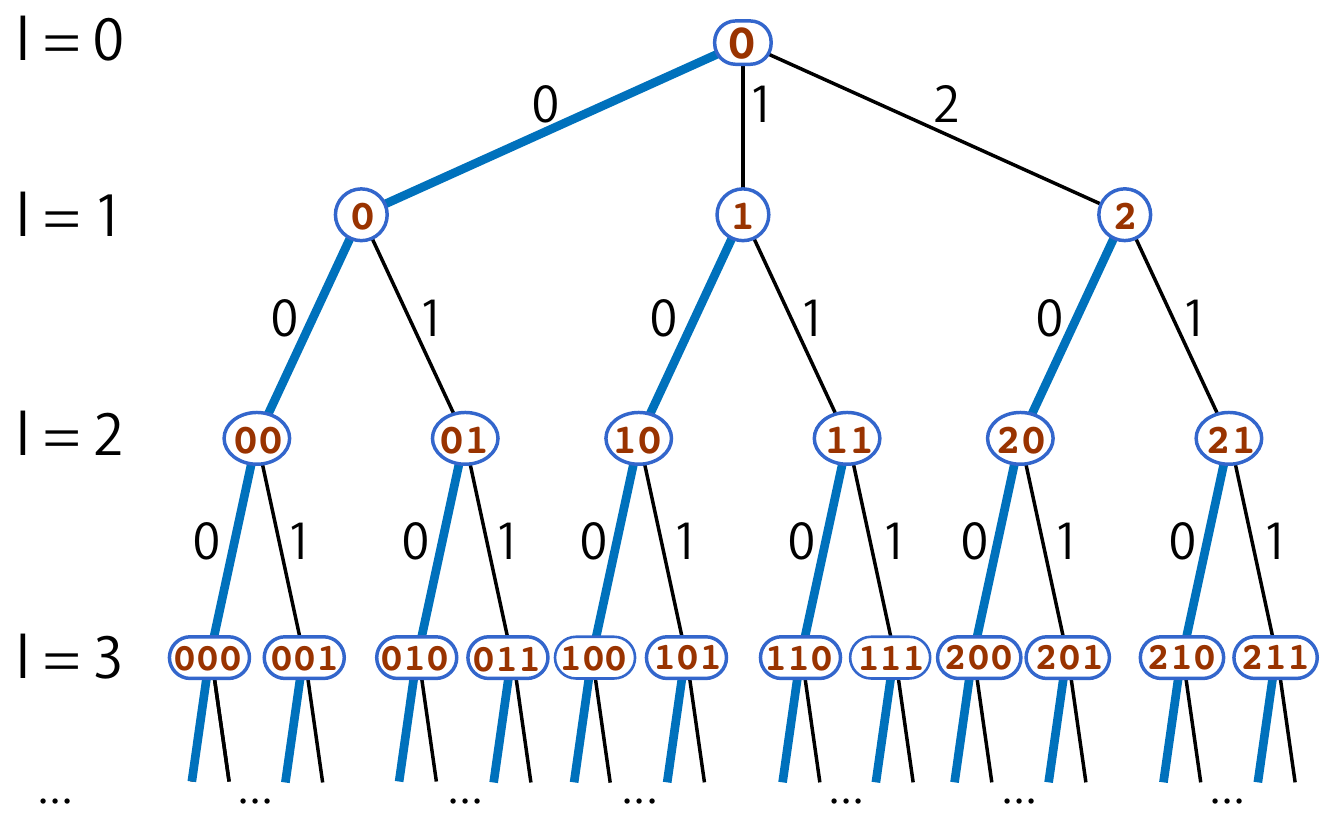}
\caption{(Color online) Site labelings on the $z=3$ Bethe lattice, in both
the real space $r$- and its dual $q$-coordination.  $l$ indicates the layers, and ``0", ``1", and ``2" on the links label different branches, the combination of which uniquely determines a path which can be further used to mark the sites. Thick blue lines indicate the effective 1D paths along which magnons propagate in the $q$-space.}
\label{Fig:Label_BL}
\end{figure}

Assuming all spins are up polarized in the ground state, we exploit the
Holstein-Primakoff (HP) transformation and take the leading approximation
for the spin operators, $S^{+}=\sqrt{2S}a$, $S^{-}=\sqrt{2S}a^{\dagger }$
and $S^{z}=S-a^{\dagger }a$, where $a$, $a^{\dagger }$ are boson
annihilation and creation operators, respectively. Under the linear
spin-wave approximation, the Hamiltonian becomes
\begin{equation}
H=(-zS\Delta
+h)\sum_{l,R_{l}}n_{R_{l}}-S\sum_{R_{l}}\sum_{r_{l+1}}(a_{R_{l}}^{\dagger
}a_{R_{l}r_{l+1}}+\mathrm{h.c.}),  \label{Eq:HPT}
\end{equation}%
where $n_{R_{l}}=a_{R_{l}}^{\dagger }a_{R_{l}}$ is the particle number
operator, and a constant energy term is dropped for the sake of simplicity.

To diagonalize the Hamiltonian (\ref{Eq:HPT}), we first take the
following multidimensional discrete Fourier transformation
\begin{equation}
a_{Q_{l}}=\sum_{R_{l}}\left( \prod_{\lambda =1}^{l}U_{q_{\lambda
},r_{\lambda }}\right) \,a_{R_{l}},  \label{Eq:QTrans}
\end{equation}%
where $Q_{l}=\left( q_{1}q_{2}...q_{l}\right) $ are the \textquotedblleft
quasi-momenta\textquotedblright\ which are dual to $R_{l}$, and $U$ is a unitary
matrix
\begin{equation}
U_{q_{\lambda },r_{\lambda }}=\frac{1}{\sqrt{\tilde{z}}}e^{2\pi \mathrm{i}%
\cdot q_{\lambda }r_{\lambda }/\tilde{z}}.
\end{equation}%
Through this transformation, we generate a Bethe lattice with exactly the
same geometry in the $q$-space, i.e., $q_{l}\in \{0,1,...,\tilde{z}-1\}$.

Under the above transformation, the hopping term in Eq.~(\ref{Eq:HPT})
becomes
\begin{equation}
\sum_{R_{l},r_{l+1}}a_{R_{l}}^{\dagger }a_{R_{l}r_{l+1}}=\sqrt{\tilde{z}}%
\sum_{Q_{l}}a_{Q_{l}}^{\dagger }a_{Q_{l},0},  \label{Eq:HopTerm}
\end{equation}%
where $\left( Q_{l},0\right) $ denotes a site on the $(l+1)$-th shell, next
to site $Q_{l}$ and with $q_{l+1}=0$. On the other hand, the occupation
number term remains as a number operator in the $q$-space
\begin{equation}
\sum_{R_{l}}a_{R_{l}}^{\dagger }a_{R_{l}}=\sum_{Q_{l}}a_{Q_{l}}^{\dagger
}a_{Q_{l}}.  \label{Eq:Occ}
\end{equation}%
More details of the transformation can be found in App.~\ref{App:rqtrans}.

Equation~(\ref{Eq:HopTerm}) with its hermitian conjugate implies that a
boson on the site $Q_{l}$ can only hop to site $\left( Q_{l},0\right)$, and vice versa.
$\left( Q_{l},0\right)$ indicate a site on $(l+1)$-th shell, 
which is on the same branch of $Q_{l}$ and further has $q_{l+1}=0$.
Therefore, for any given site $Q_{l}$ with $q_{l}\neq 0$, there exists a
one-dimensional path, consisting of $Q_{l}$, $\left( Q_{l},0\right) $, 
$\left( Q_{l},0,0\right) $, $...$, etc. 

The Hamiltonian on this half-infinite one-dimensional chain is tridiagonal, i.e.,
\begin{equation}
H_{\text{FM}}=\sum_{i=0}^{\infty }(-zS\Delta +h)b_{i}^{\dagger }b_{i}-S\sqrt{%
\tilde{z}}(b_{i}^{\dagger }b_{i+1}+\mathrm{h.c.}),  \label{Eq:HalfChain}
\end{equation}
where $b_{i}=a_{Q_{l}\underbrace{00...0}_{i}}$, and $b_{0}=a_{Q_{l}}$ is the
starting node from which the one-dimensional path, depicted by a thick line
in Fig.~\ref{Fig:Label_BL}, is defined.

\begin{figure}[!tbp]
\includegraphics[width=1\linewidth]{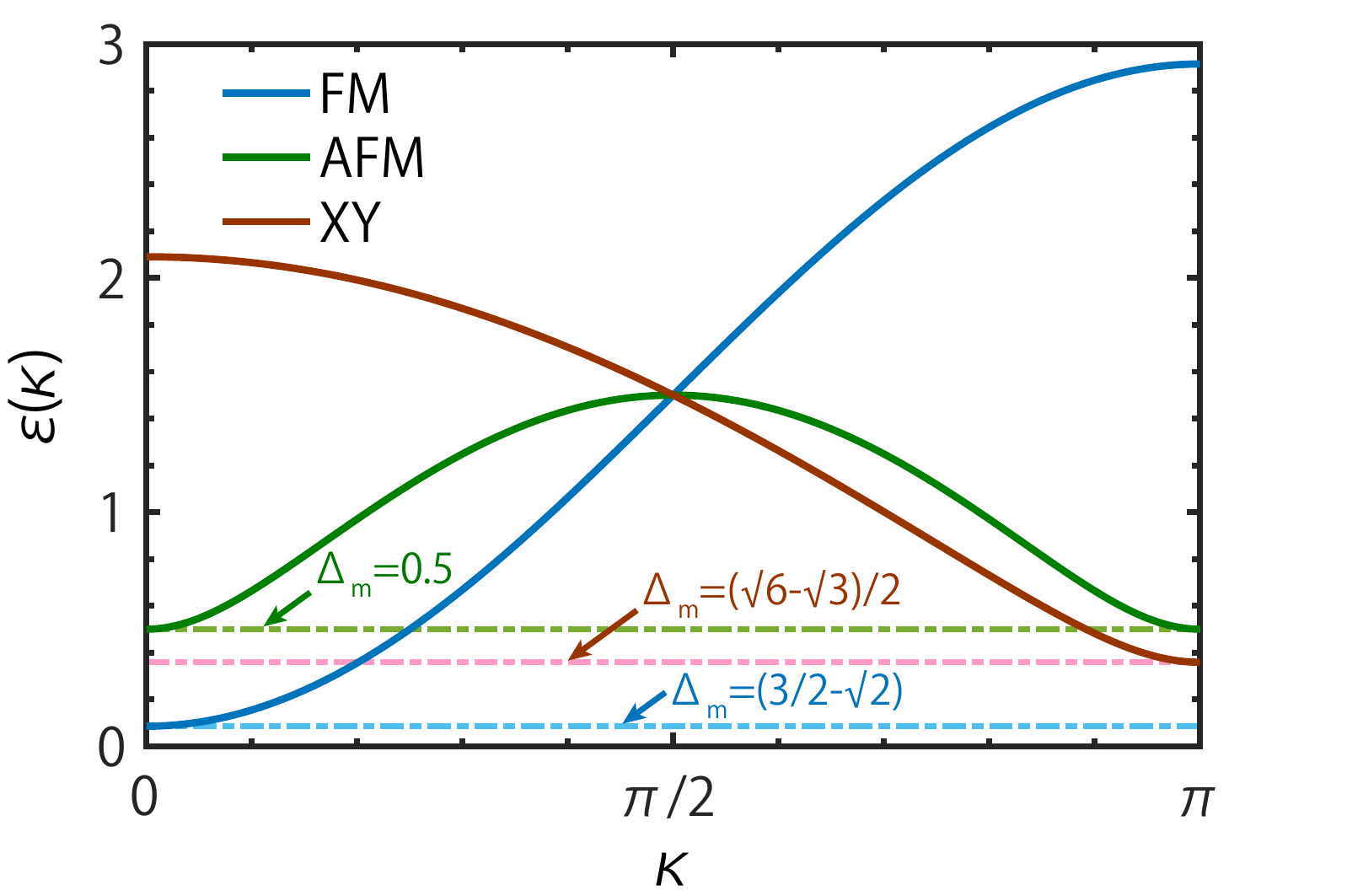}
\caption{(Color online) Magnon bands of the FM, AF, and XY Heisenberg models
on the $z=3$ Bethe lattice, where the horizontal dashed lines indicate the
magnon gap values $\Delta_m$.}
\label{Fig:MagnonBand}
\end{figure}

Secondly, we solve the above chain Hamiltonian 
via a conventional Fourier transformation, 
and the spin wave energy spectrum is found to be
\begin{equation}
\varepsilon (\kappa)=-zS\Delta +h-2S\sqrt{z-1}\cos \kappa,
\label{Eq:FMSpec}
\end{equation}
where $\kappa \in [ 0,\pi ]$ is the momentum along the ``effective" 
chains on the $q$-space Bethe lattice. From Eq.~(\ref{Eq:FMSpec}), 
as well as the plot in Fig.~\ref{Fig:MagnonBand}, we find the
magnon excitation energy gap as
\begin{equation}
\Delta _{m}=-zS\Delta +h-2S\sqrt{z-1}.  
\label{Eq:MagGap}
\end{equation}

The single magnon state is an eigenstate of the original Heisenberg FM
model, and thus the magnon gap $\Delta _{m}$ obtained in Eq.~(\ref{Eq:MagGap}) constitutes an
upper bound of the true excitation gap of the system. 
Matter of fact, as shown in
Tab.~\ref{Tab:MGap}, the energy gap obtained with this equation, $\Delta
_{m}=3/2-\sqrt{2}\simeq 0.0858$, is very close to the value estimated from
the TTN calculation. In addition, from Eq.~(\ref{Eq:MagGap}) 
one can see clearly that $\Delta _{m}$ increases linearly with $h$, 
in excellent consistency with the TTN results (see also Tab.~\ref{Tab:MGap}).

From the above discussion, we notice that each branch of magnon excitations
is confined to a one-dimensional path which is formed by a symmetrized ($q=0$%
) superposition of $\tilde{z}$ real-space sites. This symmetry has already
been exploited in some previous works 
\cite{Brinkman1970,Lepetit1999,Mahan2001,Olga2016}, 
however, the $r$-$q$ transformation we introduce 
here works in a more generic way.

\subsection{Antiferromagnetic Heisenberg model}

For the AF model with $\Delta \geq 1$, the linear spin wave expansion can be
similarly done by explicitly considering the spin orientations on the two
sublattices. We start from the configuration that all the spins are upward
aligned on the $A$-sublattice and downward aligned on the $B$-sublattice. 
Accordingly, we take a two-sublattice HP transformation, i.e., $S_{i}^{+}=\sqrt{2S}a_{i}$,
$S_{i}^{-}=\sqrt{2S}a_{i}^{\dagger }$ and $S_{i}^{z}=S-a_{i}^{\dagger }a_{i}$%
, for $i\in A$, and $S_{j}^{+}=\sqrt{2S}a_{j}^{\dagger }$, $S_{j}^{-}=\sqrt{%
2S}a_{j}$ and $S_{j}^{z}=a_{j}^{\dagger }a_{j}-S$ for $j\in B$. 
Therefore, the Hamiltonian under the linear spin wave approximation can be written as
\begin{equation}
H=zS\Delta
\sum_{l,R_{l}}n_{R_{l}}+S\sum_{l,R_{l}}%
\sum_{r_{l+1}}(a_{R_{l}}a_{R_{l}r_{l+1}}+a_{R_{l}}^{\dagger
}a_{R_{l}r_{l+1}}^{\dagger }).  \label{Eq:HPT2}
\end{equation}%
A constant term is again omitted in obtaining the above expression.

To cope with the two-sublattice structure, we need to adopt different
transformations on diffferent sublattices, namely taking $\tilde{U}%
_{q,r}=U_{q,r}$ on the $A$-sublattice and $\tilde{U}_{q,r}=U_{q,r}^{\dagger }
$ on the $B$-sublattice. Under this transformation, the pair annihilation
term becomes (App. \ref{App:rqtrans})
\begin{equation}
\sum_{R_{l},r_{l+1}}a_{R_{l}}a_{R_{l}r_{l+1}}=%
\sum_{Q_{l}}a_{Q_{l}}a_{Q_{l},0}.  \label{Eq:Pair}
\end{equation}%
The above Hamiltonian can then be effectively represented as a direct sum of
the infinite-many one-dimensional models defined by the model
\begin{equation}
H_{\mathrm{AF}}=\sum_{i=0}^{\infty }S[z\Delta \,b_{i}^{\dagger }b_{i}+\sqrt{%
\tilde{z}}(b_{i}b_{i+1}+b_{i}^{\dagger }b_{i+1}^{\dagger})].
\label{Eq:HalfChain2}
\end{equation}%
Employing the standard Bogoliubov transformation, the energy spectrum of $H_{\mathrm{AFC}}
$ is found to be (and plotted in Fig.~\ref{Fig:MagnonBand})%
\begin{equation}
\varepsilon (\kappa )=S\sqrt{(z\Delta )^{2}-4(z-1){\cos }^{2}\kappa },\text{
\ \ \ } \kappa \in \lbrack 0,\pi ].
\end{equation}%
The energy gap is, $\Delta _{m}=S(z-2)$, for $z\geq 3$. Since the
single-magnon excitation state is not an eigenstate of the original AF
model, $\Delta _{m}$ may not be an upper bound of the true excitation gap.
Nonetheless, from Tab.~\ref{Tab:MGap} we observe that
 $\Delta _{m}$ still provides a quite good estimate of the magnon gap.

\subsection{The XY model}

The magnon bands of FM and AF cases become unstable when $|\Delta |<1$ (more
precisely, when $|\Delta |<2\sqrt{z-1}/z$). To perform a linear spin-wave
analysis for the XY phase, we needs to start from a classical state in which
the spins are ordered on the XY\ plane.

Below, for the sake of simplicity, we consider only the case $\Delta =0$. The
Hamiltonian can be equivalently written as
\begin{equation}
H=\sum_{\langle i,j\rangle } S_{i}^{z}S_{j}^{z}+S_{i}^{x}S_{j}^{x}.
\end{equation}%
Under the linear spin-wave approximation, it becomes
\begin{equation}
H=zS\sum_{l,R_{l}}n_{R_{l}}+\frac{S}{2}\sum_{l,R_{l}}%
\sum_{r_{l+1}}(a_{R_{l}}+a_{R_{l}}^{\dagger
})(a_{R_{l}r_{l+1}}+a_{R_{l}r_{l+1}}^{\dagger }).  \label{Eq:BosonXY}
\end{equation}

To solve this problem, we use the following $r$-$q$ transformation matrix
\begin{equation}
U_{q,r}=%
\begin{cases}
1/\sqrt{\tilde{z}}, & q=0, \\
\sqrt{\frac{2}{\tilde{z}}}\cos [\frac{\pi }{\tilde{z}}(r+\frac{1}{2})q], &
q=1,2,...,\tilde{z}-1.%
\end{cases}%
\end{equation}%
This particular choice ensures $U_{q,r}$ to be a real 
orthogonal matrix with the property
\begin{equation*}
\sum_{r}U_{q,r}=\sqrt{\tilde{z}}\delta _{q,0}.
\end{equation*}%
From this, again we obtain an effective one-dimensional boson model
\begin{equation}
H_{\mathrm{XY}}=\sum_{i=0}^{\infty }S[zb_{i}^{\dagger }b_{i}+\frac{\sqrt{%
\tilde{z}}}{2}(b_{i}+b_{i}^{\dagger })(b_{i+1}+b_{i+1}^{\dagger })].
\label{Eq:XYboson}
\end{equation}%
Diagonalizing this Hamiltonian using the Bogoliubov transformation (App.~\ref%
{App:Bogoliubov}), the magnon excitation energy is found to be (Fig.~\ref%
{Fig:MagnonBand})
\begin{equation}
\varepsilon (\kappa )=S\sqrt{z(z+2\sqrt{z-1}\cos \kappa )},\text{ \ \ \ \ }%
\kappa \in \lbrack 0,\pi ].  \label{Eq:BandXY}
\end{equation}%
The excitation gap is $\Delta _{m}=S\sqrt{z}(\sqrt{z-1}-1)$, which
is $(\sqrt{6}-\sqrt{3})/2\approx 0.3587$ for $S=1/2$ and $z=3$, close to the
numerical results, $0.41(1)$, shown in Tab.~\ref{Tab:MGap}.

\section{Summary}

\label{Sec:Summary}

We have investigated the Heisenberg XXZ model using both the canonical TTN
and linear spin-wave theory on the $z=3$ Bethe lattice. Through efficient
and accurate tensor network simulations, 
we have obtained the finite-temperature phase diagram of
the model. The system undergoes a second-order phase transition
at finite temperature,
and three kinds of magnetic 
ordered phases are uncovered in low $T$. 
The correlation lengths, as well as bipartite entanglements, 
though exhibiting their maximal values at the transition temperature $T_c$, 
are found to be always finite on the Bethe lattice. 
Correspondingly, the low-lying excitations are revealed, through both numerical TTN and analytical spin wave calculations, to be gapped,
even in the parameter range where the system breaks spontaneously the continuous symmetries.  

Therefore, although the conventional gapless Goldstone modes are absent, 
quasi-Goldstone transverse 
fluctuation modes have been observed in the Bethe lattice XXZ model. 
When the system spontaneously breaks the continuous symmetries, 
the transverse correlation lengths reach the \textquotedblleft critical\textquotedblright\
value, $\xi _{c}=1/\ln {(z-1)}$, and remain there for $T\leq T_c$. Here $\xi _{c}$ is
the maximal correlation length that is allowed on the Bethe lattice.

The results obtained on the Bethe-like lattices can be used to understand physical properties of quantum lattice models on the honeycomb, square, or other regular lattices. In addition, the canonical TTN method we proposed works very generally, and it can be applied to other fundamental quantum many-body models, such as the frustrated Heisenberg and Hubbard models defined on the Bethe-like lattices, etc.

\section{acknowledgements}

This work was supported by the National Natural Science Foundation of China
(11834014, 11888101), the National R\&D Program of China (2017YFA0302900).
WL is indebted to Andreas Weichselbaum for helpful discussions. DWQ and WL would like to
thank Jan von Delft for hospitality during a visit to LMU Munich, where part
of this work was performed.

\begin{appendix}

\setcounter{equation}{0}
\setcounter{figure}{0}
\setcounter{table}{0}


\renewcommand{\theequation}{A\arabic{equation}}
\renewcommand{\thefigure}{A\arabic{figure}}
\renewcommand{\bibnumfmt}[1]{[#1]}
\renewcommand{\citenumfont}[1]{#1}

\section{Tensor update on the Bethe lattice}
\label{App:BetheAlg}

\subsection{Bethe lattice update}
\label{App:SimpleU}
\begin{figure}[!tbp]
\includegraphics[width=1\linewidth]{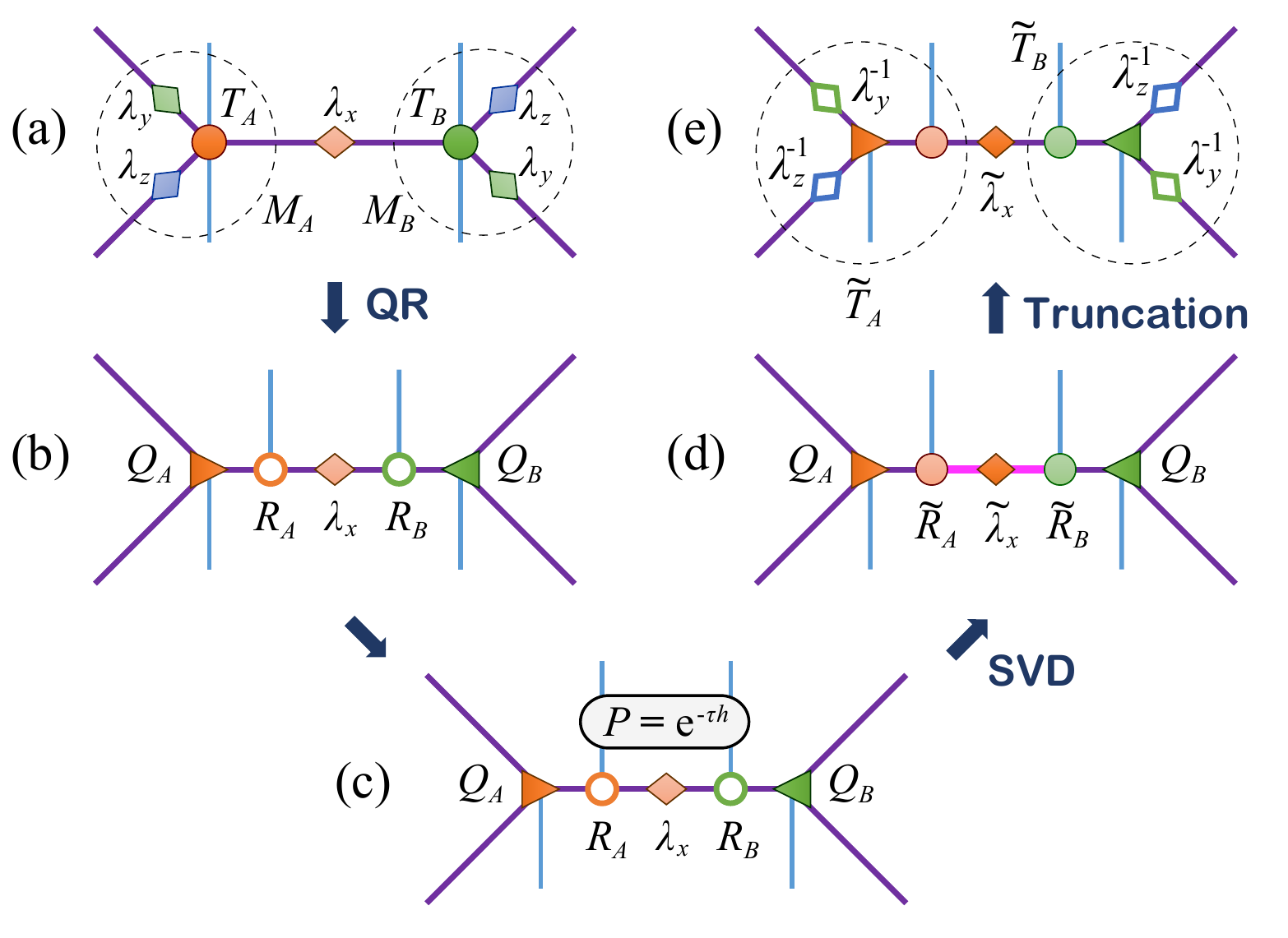}
  \caption{(Color online) A single step of imaginary-time evolution on an $x$-bond, see Sec.~\ref{App:SimpleU} for details.
}
\label{Fig:Proj_BL}
\end{figure}

  An imaginary time evolution can be employed to cool down the TTN density operator on the Bethe lattice, where the local tensors are updated via a simple scheme. The details are illustrated in Fig.~\ref{Fig:Proj_BL} and elaborated below. We take the the $x$-bond update as an example, and the successive projection procedures on the other two, i.e., $y$- and $z$-bonds, can be accomplished similarly. The three projection substeps on different bonds constitute a full Trotter step of small imaginary-time slice $\tau$.

  (a) Absorb the environment matrices $\lambda_y$ and $\lambda_z$ to the $T$ tensors and construct the $W$ tensor [Fig. \ref{Fig:Proj_BL}(a)]
\begin{equation}\label{Eq:Proj_BL1}
  (W_\alpha)_{xyz}^{\sigma_\alpha \sigma_\alpha^\prime} = (T_{\alpha})_{xyz}^{\sigma_\alpha \sigma_\alpha^\prime} \lambda_y \lambda_z,
\end{equation}
  where $\alpha = A,B$.
	
  (b) Perform a QR decomposition of $W_{\alpha}$ [Fig. \ref{Fig:Proj_BL}(b)]
\begin{equation}\label{Eq:Proj_BL2}
  (W_\alpha)_{xyz}^{\sigma_\alpha \sigma_\alpha^\prime} = \sum_{x^\prime} (Q_\alpha)_{x^\prime yz}^{\sigma_\alpha^\prime} (R_\alpha)_{x^\prime x}^{\sigma_\alpha},
\end{equation}
  which splits off the upper physical indices $\sigma_\alpha$ into the $R$ tensors.
	
  (c) Construct a base tensor $B$ by combining $\lambda_x$ with the two adjacent tensors $R_A$ and $R_B$
\begin{equation}\label{Eq:Proj_BL3}
  B_{x_A^\prime x_B^\prime}^{\sigma_A \sigma_B} = \sum_x (R_A)_{x_A^\prime x}^{\sigma_A}  \lambda_x (R_B)_{x_B^\prime x}^{\sigma_B},
\end{equation}
  and apply the two-site imaginary-time evolution gate $P=e^{-\tau h_x}$ onto the base tensor [Fig.~\ref{Fig:Proj_BL}(c)]
\begin{equation}\label{Eq:Proj_BL4}
  \tilde{B}_{x_A^\prime x_B^\prime}^{\sigma_A \sigma_B} = \sum_{\sigma_A^\prime,\sigma_B^\prime} P_{\sigma_A^\prime \sigma_B^\prime}^{\sigma_A \sigma_B} B_{x_A^\prime x_B^\prime}^{\sigma_A^\prime \sigma_B^\prime}.
\end{equation}
	  	
  (d) Reshape $\tilde{B}$ into a matrix by grouping together $\sigma_A$ with $x_A^\prime$ as a single index, and $\sigma_B$ with $x_B^\prime$ into another. Then we perform a matrix SVD
\begin{equation}\label{Eq:Proj_BL5}
  (\tilde{B})_{x_A^\prime x_B^\prime}^{\sigma_A \sigma_B} = \sum_x (\tilde{R}_A)_{x_A^\prime x}^{\sigma_A}  \tilde{\lambda}_x (\tilde{R}_B)_{x_B^\prime x}^{\sigma_B},
\end{equation}
  as shown in Fig.~\ref{Fig:Proj_BL}(d). Note that the matrix dimension of $\tilde{\lambda}_x$ is enlarged by $d^2$ times after the bond evolution, which needs to be truncated by retaining only the largest $D$ singular values and corresponding bond bases. After this proper truncation of bond states, we update the tensors $\tilde{R}_a$, $\tilde{R}_b$ and $\tilde{\lambda}_x$ accordingly.

(e) Combine $\tilde{R}$ to the corresponding $Q$ tensors, and spit off the $\lambda_{y(z)}$ matrices, that have been absorbed into $Q$ in steps (a,b), by multiplying their inverse matrices $\lambda^{-1}_{y(z)}$ to $Q$. Thus we obtain the updated $\tilde{T}$ tensors [Fig. \ref{Fig:Proj_BL}(e)]
\begin{equation}\label{Eq:Proj_BL6}
(\tilde{T}_\alpha)_{xyz}^{\sigma_\alpha \sigma_\alpha^\prime} = \sum_{x^\prime} \lambda_y^{-1} \lambda_z^{-1} (Q_\alpha)_{x^\prime yz}^{\sigma_\alpha^\prime} (\tilde{R}_\alpha)_{x^\prime x}^{\sigma_\alpha},
\end{equation}
  where again $\alpha=A, B$ denoting two sublattices.
	
  In the groundstate optimization \cite{Li2012}, the above procedure suffices to produce accurate results. The truncation errors do not accumulate, and $\tau$ can be tuned to a sufficiently small value, say, $\tau \sim10^{-4}$ in the end of projections. However, in the finite-temperature simulations, now the truncation errors accumulate, we therefore need to carefully optimize the truncations in every single step to improve the overall performance. Note that $e^{-\tau h_\zeta^i}$ ($\zeta = x,y,z$) is not unitary and breaks the orthogonality of bond basis, and thus TTN deviates the canonical form
after each step of imaginary-time evolution. Therefore, a canonicalization procedure of the TTN is needed to restore the orthogonality of bond bases and optimize the truncations.

\subsection{Canonicalization procedure of TTN}
\label{App:CanonBethe}

\begin{figure}[!tbp]
\includegraphics[width=1\linewidth]{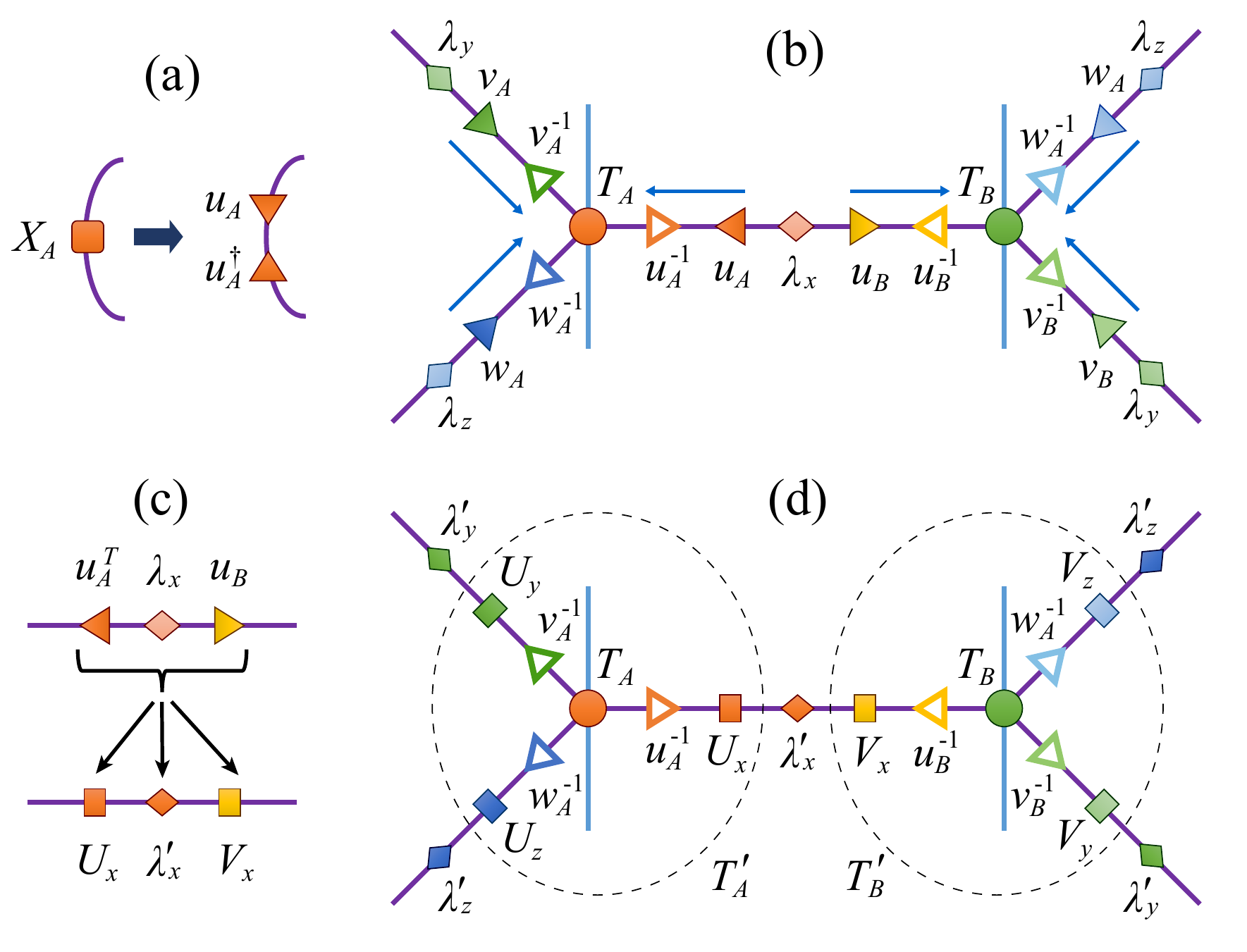}
  \caption{(Color online) Canonicalization procedure of the TTN, where
  the $x$-bond procedure is taken as an example in (a) and (c). See Sec.~\ref{App:CanonBethe} for details.
}
\label{Fig:CanT_BL}
\end{figure}

Following a similar line of standard procedure developed in the matrix product \cite{Vidal2007,OrusVidal2008,Orus2014} as well as tensor product states \cite{Ran.s+:2012:Super-orthogonalization}, we present below the canonicalization of TTN on the Bethe lattice.

(i) As shown in Fig.~\ref{Fig:CanT_BL}(a), we decompose the dominant eigenvectors of the transfer tensors as
\begin{equation}
\begin{split}
& (X_\alpha)_{x,x'} = \sum_{x''} (u_\alpha)_{x,x''} (u_\alpha^\dagger)_{x'',x'}, \\
& (Y_\alpha)_{y,y'} = \sum_{y''} (v_\alpha)_{y,y''} (v_\alpha^\dagger)_{y'',y'}, \\
& (Z_\alpha)_{z,z'} =\sum_{z''} (w_\alpha)_{z,z''} (w_\alpha^\dagger)_{z'',z'},
\end{split}
\end{equation}
where $\alpha=A,B$. Since in practice $X_\alpha,Y_\alpha$, and $Z_\alpha$ matrices are symmetric, the decomposition can be done via the eigenvalue or the Cholesky decomposition.

(ii) Insert pairs of reciprocal matrices, $u_\alpha u_\alpha^{-1}$, $v_\alpha v_\alpha^{-1}$ and $w_\alpha w_\alpha^{-1}$, to the three geometrical bonds, as shown in Fig. \ref{Fig:CanT_BL}(b). Order of the matrix multiplication has also been specified by the arrows, e.g., we contract the first index of $u_{A}$ with $\lambda_x$ and the second index of $u_{A}^{-1}$ with $T_A$.
	
(iii) Now we perform the bond update by combining the bond matrices and then perform a SVD. As shown in Fig. \ref{Fig:CanT_BL}(c), we take the $x$-bond as an example, i.e.,
\begin{equation}
u_A^T \lambda_x u_B = U_x \lambda_x^\prime V_x.
\end{equation}
$U_x$ and $V_x$  are unitary matrices, and
$\lambda'_x$ is used to update the bond diagonal matrix.
	
(iv) As shown in Fig. \ref{Fig:CanT_BL}(d), we absorb $u^{-1}$, $v^{-1}$ and $w^{-1}$ matrices, as well as the adjacent unitary matrices $U$ and $V$, into the $T_\alpha$ tensors, and update $T_A^\prime$ and $T_B^\prime$.

 After the above procedure on the $x$ bond (and simultaneously on the $y$ and $z$ bonds), the updated tensors $T_\alpha^\prime$ satisfy the canonical conditions [see Eq.~(\ref{Eq:CanonCond}) of the main text], and the dominating eigenvectors $X_\alpha, Y_\alpha, Z_\alpha$ are now gauged into identities.

\subsection{Simple vs. canonical schemes}
\label{App:SvsC}

\begin{figure}[!tbp]
\includegraphics[width=0.9\linewidth]{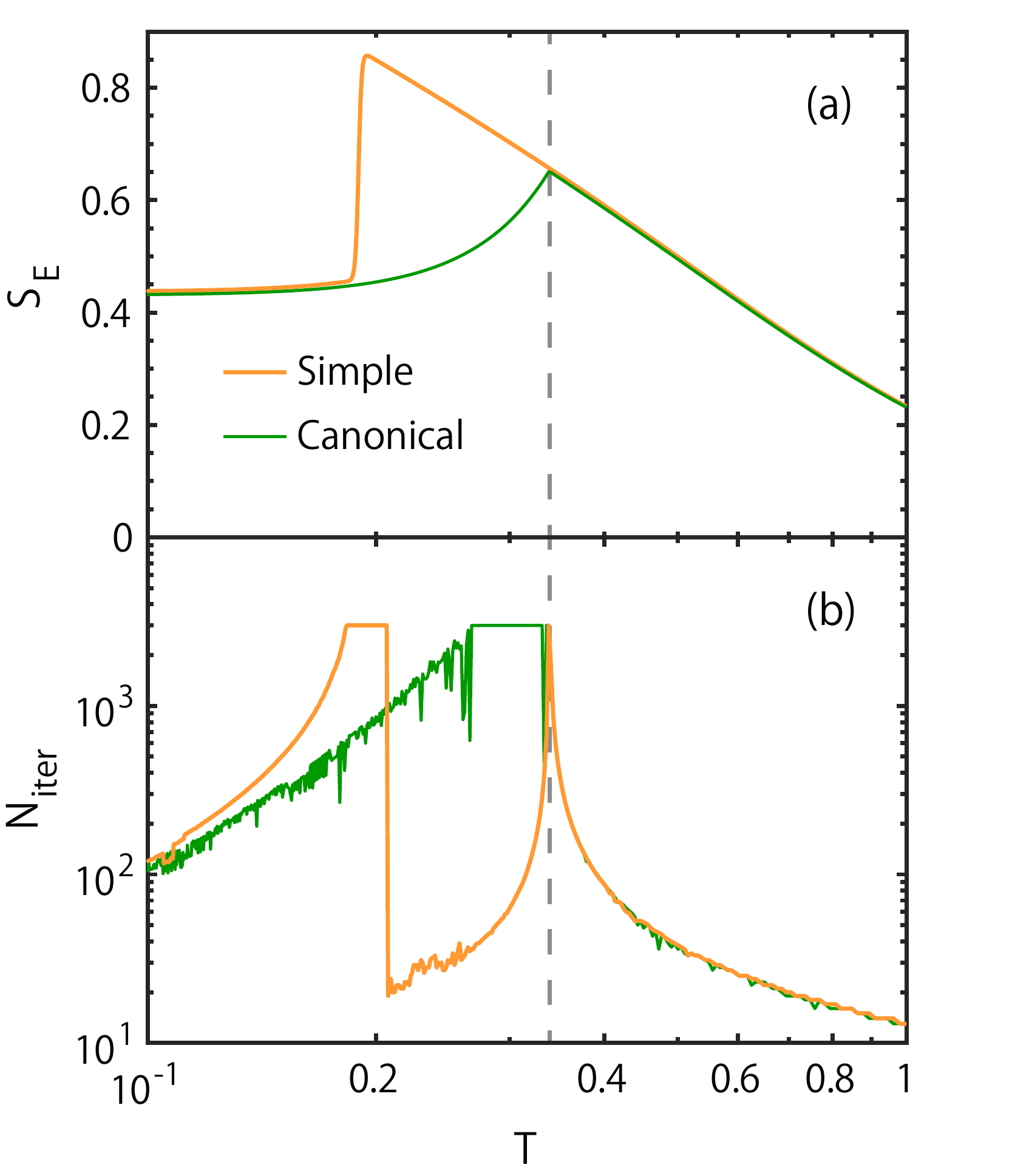}
\caption{(Color online) Comparisons between the simple and canonical update optimization schemes on (a) the entanglement entropy $S_E$ and  (b) iteration number $N_{\rm{iter}}$ for $\Delta=1$. The dashed gray line denotes the critical point $T_c$ for $\Delta=1$, and the convergence tolerant for the iterations is $10^{-7}$, with the maximal iteration number restricted up to 3000 in practice.}
\label{Fig:SvsC}
\end{figure}

  Now we provide some numerical benchmarks of the simple and canonical update schemes, showing the advantage of the latter in both the accuracy and robustness. Here by canonical scheme we mean the combined procedure using techniques introduced in both Secs.~\ref{App:SimpleU} and \ref{App:CanonBethe}; while by simple scheme we mean a poorman's approach where the canonicalization operations in Sec.~\ref{App:CanonBethe} are skipped.

  We consider the Heisenberg model ($\Delta=1$), and compare the entanglement entropy $S_E$ and the number of iterations $N_{\rm{iter}}$ required to reach a convergence in determining the dominant eigenvectors in Eq.~(\ref{Eq:Iter_BL}) of the main text. Although the entanglement entropy $S_E$ defined in Eq.~(\ref{Eq:Se}) is rigorously defined only for canonical TTN, we can nevertheless take ``$S_E$" from the simple scheme as a measurement of entanglement for comparisons.

  As seen in Fig.~\ref{Fig:SvsC}(a), the two $S_E$ curves almost coincide for $T > T_c$. However, their behaviors start to differ at the critical temperature $T_c$. The curve of the canonical scheme shows a cusp at the critical temperature and slowly converges to a smaller zero-temperature entanglement value, while that of the simple scheme still rises smoothly until collapsing at certain lower temperature below $T_c$. After this ``jump", the simple scheme curve lies almost on top of the canonical curve, and both converge to the $T=0$ entanglement value in nearly the same rate.

  Correspondingly, as shown in Fig.~\ref{Fig:SvsC}(b), $N_{\rm{iter}}$ peaks at $T_c$ in the canonical scheme, and it peaks both at $T_c$ and the ``jump" point at a lower $T$ in the simple scheme curve. The second peak in $N_{\rm{iter}}$ in the simple scheme belongs to an numerical ``artifact", since no phase transition really takes place there.

  Apart from the ``artifact'' in $S_E$, the more severely accumulated errors in the simple scheme, as the temperature lows down, may cause other problems. In practice, sometimes the simple scheme is found to generate ``wrong" metastable thermodynamic results, e.g, magnetic moments, at low temperatures $T<T_c$.

  To conclude, the canonical scheme turns out to be more accurate and robust, and it is thus mostly adopted in our practical simulations.
	
\section{Classical Ising model on the Bethe lattice}
\label{App:Ising}
In this appendix, we provide the rigorous solution of the classical Ising model on the Bethe lattice via transfer tensor techniques.  The TTN algorithms introduced in Sec.~\ref{Sec:MoMe} can also be employed to compute this classical  model, and the comparisons between the numerical and rigorous results thus provide a first benchmark of the TTN algorithm.

  The Hamiltonian (energy) of the classical Ising model reads $H=-J\sum_{\langle i,j \rangle} \sigma_i \sigma_j$, where $J=1$ is the energy scale, $\langle i,j \rangle$ means NN sites on the Bethe lattice, and $\sigma=\pm1$ denotes the classical Ising variables.

  This  Ising model can be solved exactly through a number of essentially equivalent methods, including the self-similarity \cite{Baxter2013,Ostilli2012} and cavity approaches \cite{Mezard2001,Ostilli2012}, etc. By solving the generalized eigenvalue problem, we find the dominant eigenvectors of the transfer tensor $\mathcal{T}$, from which we further obtain the exact expression of thermal quantities.

\begin{figure}[!tbp]
\includegraphics[width=0.95\linewidth]{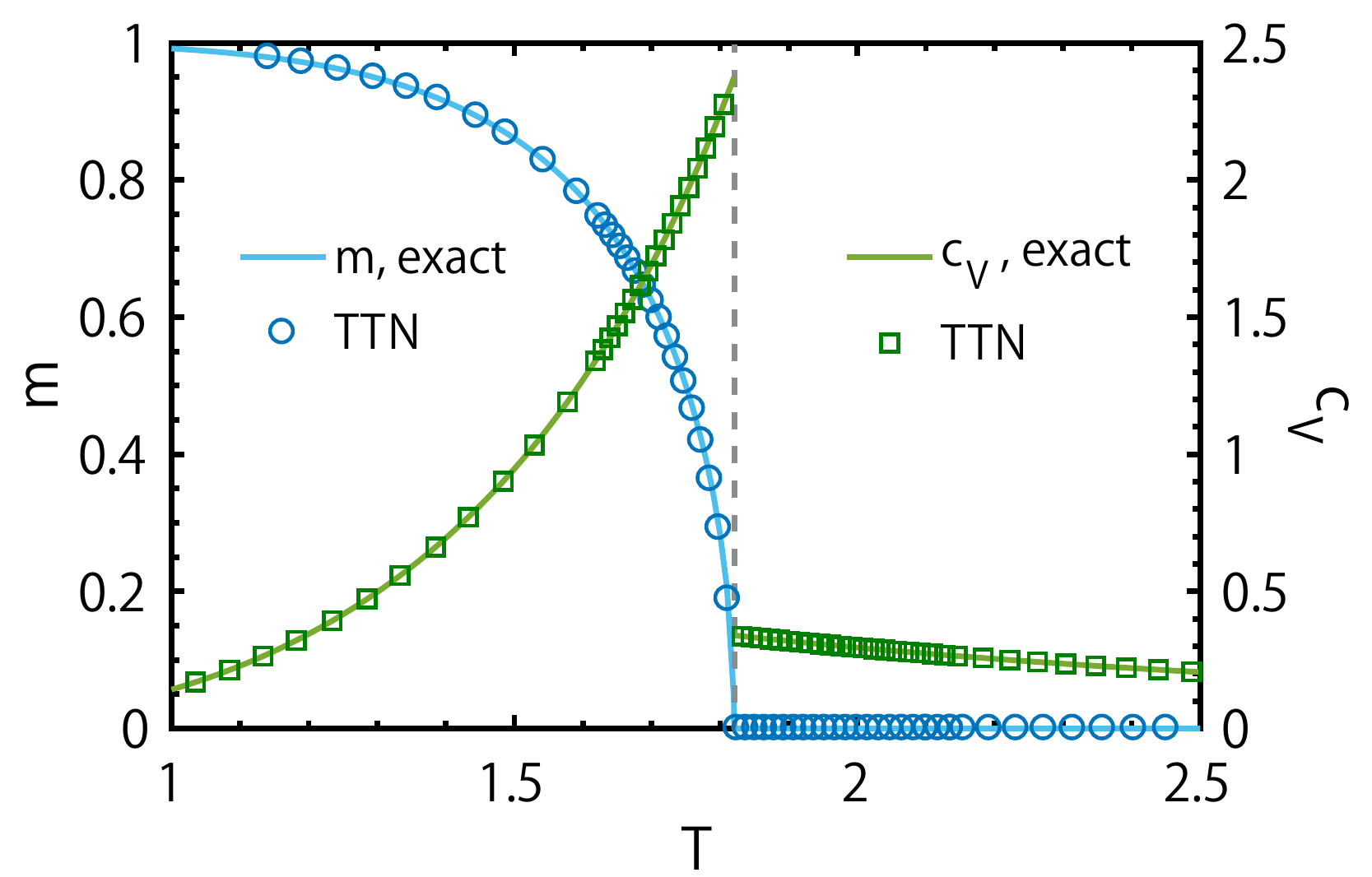}
\caption{(Color online) Comparisons between the TTN calculations and the exact solution of the classical Ising model on the Bethe lattice, where a perfect agreement is seen in both the magnetic moment ($m = \lvert\langle\sigma\rangle\rvert$) and the specific heat $c_V$ curves. The vertical dashed line denotes the critical temperature $T_c = 2/\ln(3)$.
}
\label{Fig:Ising_BL}
\end{figure}

Firstly, we rewrite the partition function as a TTN, i.e.,
\begin{equation}
\label{Eq:BetheIsing}
\mathcal{Z} = \sum_{\{\sigma\}} e^{\beta\sigma_1\sigma_2} \mathcal{T}^{\sigma_{11}\sigma_{12}}_{\sigma_1} \mathcal{T}^{\sigma_{21}\sigma_{22}}_{\sigma_2} \mathcal{T}^{\sigma_{111}\sigma_{112}}_{\sigma_{11}} \mathcal{T}^{\sigma_{121}\sigma_{122}}_{\sigma_{12}}...
\end{equation}
where $\mathcal{T}$ is a transfer tensor, and $\sigma_{1(2)}$ sits in the two central sites in the bond-centered  [see, e.g., Fig.~\ref{Fig:Tree}(b)]. The $r_1r_2...r_l$ labelling in $\sigma_{r_1r_2...r_l}$ follows the convention shown in Fig.~\ref{Fig:Label_BL}, but $r_i$ starts from 1 instead of 0 here.

Efficient contractions of this infinite TTN can be implemented by finding the dominant (generalized) eigenvectors $V_\sigma$ of the transfer tensor $\mathcal{T}^{\sigma_{1}\sigma_{2}}_{\sigma_0} = e^{\beta\sigma_0(\sigma_1+\sigma_2)}$, i.e.,
\begin{equation}
\label{Eq:IsingV}
\sum_{\sigma_1,\sigma_2} \mathcal{T}^{\sigma_{1}\sigma_{2}}_{\sigma_0} V_{\sigma_1} V_{\sigma_2} = \eta V_{\sigma_0}
\end{equation}
with the (generalized) eigenvalue $\eta$. By writing it down explicitly, we have
\begin{equation}
\label{Eq:Vpm}
\begin{split}
V_+^2 e^{2\beta} + V_-^2 e^{-2\beta} + 2V_+V_-  = \eta V_+, \\
V_+^2 e^{-2\beta} + V_-^2 e^{2\beta} + 2V_+V_-  = \eta V_-.
\end{split}
\end{equation}

Since Eq.~(\ref{Eq:Vpm}) does not constitute a linear system of equations, one has a gauge degrees of freedom in determining the eigenvalue $\eta$ as well as the eigenvector $V_{\pm}$. Nevertheless, we can eliminate $\eta$ from the equations, and arrive at a cubic equation after some rearrangement,
\begin{equation}\label{Eq:cubic}
    (x-1)[x^2+(1-e^{2\beta})x+1]=0,
\end{equation}
where $x = \sqrt{V_+/V_-}$. Note that the eigenvalue can now be expressed as $\eta=V_+ (x/e^\beta + e^\beta/x)^2$, which can not be uniquely determined and depends on the normalization condition of $V_\sigma$ (see related discussions in Sec. \ref{SSec:ITE_BL}).

  The (unnormalized) probability distribution for two neighboring spin variables is $\rho_{\sigma_1\sigma_2} = e^{\beta\sigma_1\sigma_2}V_{\sigma_1} V_{\sigma_2}$. By summing over $\sigma_2$, we can get the distribution of a single spin $\sigma_1$, from which the magnetization can be derived as $\langle\sigma\rangle = \sum_{\sigma_1} \sigma_1\rho_{\sigma_1} / \sum_{\sigma_1}\rho_{\sigma_1} = (x^3-1)/(x^3+1)$. Note that $x$ is just the root of Eq.~(\ref{Eq:cubic}), and thus the local magnetization can be uniquely determined.

The order parameter, i.e., the spontaneous magnetization $m$, reads
\begin{equation}
\label{Eq:IsingSz_BL}
m = \langle \sigma_1 \rangle_\beta =
\begin{cases}
0, & \beta \leq \beta_c, \\
\pm\frac{\omega}{\omega-2} \sqrt{\frac{\omega-3}{\omega+1}}, & \beta>\beta_c,
\end{cases}
\end{equation}
with $\omega = e^{2\beta}$ and the critical point $\beta_c = 1/T_c = \ln(3)/2\simeq 0.5493$.

Apparently, the root $x=1$ (and thus $m=0$) corresponds to the paramagnetic solution, while the two roots of the remaining quadratic equation in Eq.~(\ref{Eq:cubic}) reflects the two-fold degenerate FM states.
The $\pm$ sign represents the spin-up and spin-down solutions, respectively, given that the discriminant $(1-e^{2\beta})^2-4 > 0$, i.e.,
$\beta>\beta_c=\ln(3)/2$.

The internal energy per bond $u_b$ can be determined from $\rho_{\sigma_1\sigma_2}$, which is
\begin{equation}
\label{Eq:IsingEb_BL}
  u_b =  -\langle \sigma_1\sigma_2 \rangle_\beta =
\begin{cases}
  \frac{1-\omega}{1+\omega}, & \beta \le \beta_c \\
  \frac{4}{(\omega^2-1)(\omega-2)}-1. & \beta>\beta_c
\end{cases}
\end{equation}
It is clear that the $u_b$ curve exhibits a singular point at the transition temperature $\beta_c$.

On the other hand, the Bethe-lattice Ising model can also be solved by the TTN techniques. By performing a decomposition following Eq.~(\ref{Eq:TS_BL}), and then an imaginary-time evolution procedure,
we obtain the thermal density matrix $\rho(\beta) = e^{\beta\sum_{\langle i,j \rangle} \sigma_i \sigma_j}$ of the classical Ising model on the Bethe lattice. On top of that, we can further calculate the thermal quantities, including the magnetization, energy, and specific heat, etc. 

  In Fig.~\ref{Fig:Ising_BL}, we compare the TTN results of the spontaneous magnetization and the specific heat to the exact solution, where excellent agreements can be observed. This can be ascribed to the absence of Trotter errors in the calculations, and also to no essential truncations in the procedure of cooling. 
  
  It is also of interest to note, in Eq.~(\ref{Eq:IsingSz_BL}), that the spontaneous magnetization $m \sim (\frac{T_c-T}{T_c})^{\kappa}$ with $\kappa=1/2$, when $T$ approaches the transition temperature $T_c$ (from low temperature side). In addition, the internal energy $u_b$ curve is continuous at $T_c$,  
while the slope $c_V \equiv \partial u_b/\partial T$ shows a discontinuity in Fig.~\ref{Fig:Ising_BL}(b), suggesting a mean-field-type critical exponent $\alpha=0$.

\begin{widetext}
\section{The $r$-$q$ transformation}
\label{App:rqtrans}
Here we provide more details on the $r$-$q$ transformations of the bilinear terms, including the hopping, on-site occupation number, pair-creation or annihilation terms, etc.

Firstly, we check that the real-space on-site term remains as on-site term in the $q$-space
\begin{equation}
\begin{split}
\sum_{R_l} a_{R_l}^\dagger a_{R_l}
& = \sum_{R_l} \sum_{Q_l,Q'_l} a^\dagger_{Q_l} \left(\prod_{\lambda=1}^l U_{q_\lambda r_\lambda} U^\dagger_{r_\lambda q'_\lambda} \right) \, a_{Q'_l} \\
& = \sum_{Q_l,Q'_l} a^\dagger_{Q_l} \left(\prod_{\lambda=1}^l \sum_{r_\lambda} U_{q_\lambda r_\lambda} U^\dagger_{r_\lambda q'_\lambda} \right) \, a_{Q'_l} \\
& = \sum_{Q_l,Q'_l} a^\dagger_{Q_l}  a_{Q'_l} \delta_{Q_l,Q'_l}
 =  \sum_{Q_l} a_{Q_l}^\dagger a_{Q_l}.
\end{split}
\end{equation}

  For the NN hopping term, we have
\begin{equation}
\begin{split}
    \sum_{R_l, r_{l+1}}  a_{R_l}^\dagger a_{R_l r_{l+1}}
    & = \sum_{R_l, r_{l+1}}\sum_{Q_l',Q_l,q_{l+1}}  a_{Q_l'}^\dagger \prod_{\lambda=1}^l U_{q'_\lambda,r_\lambda}  \prod_{\lambda=1}^{l+1} U^{\dagger}_{r_\lambda,q_\lambda}  \, a_{Q_lq_{l+1}} \\
   &   = \sum_{r_{l+1}}\sum_{Q_l',Q_l,q_{l+1}}  \delta_{Q_l',Q_l} a_{Q_l'}^\dagger  U^{\dagger}_{r_{l+1},q_{l+1}} a_{Q_lq_{l+1}} \\
    & = \sum_{Q_l,q,r}\frac{1}{\sqrt{\tilde{z}}} e^{-2\pi \mathrm{i} \cdot q r} a_{Q_l}^\dagger a_{Q_lq}
    = \sqrt{\tilde{z}} \sum_{Q_l} a_{Q_l}^\dagger a_{Q_l0},
\end{split}
\end{equation}
  where $\tilde{z} = z-1$ for $l>1$ layers and $\tilde{z}=z$ for $l=1$ one.

  In the HAF model, we have pair-creation and annihilation operators, where additional care needs to be taken of, i.e.,
\begin{equation}
a_{Q_l} =
\begin{cases}
\sum_{R_l} (\prod_{\lambda=1}^l U_{q_\lambda,r_\lambda}) \cdot a_{R_l} , & l \in \mathrm{odd}, \\
\\
\sum_{R_l}  a_{R_l} \cdot (\prod_{\lambda=1}^l U^\dagger_{r_\lambda,q_\lambda}), & l \in \mathrm{even}.
\end{cases}
\end{equation}
Therefore, the pair creation and annihilation operators can be transformed into $q$-space in a well organised way, e.g.,
\begin{equation}
\begin{split}
    & \sum_{R_l, r_{l+1}}  a_{R_l} a_{R_l r_{l+1}} \\
    &= \sum_{R_l, r_{l+1}}\sum_{Q_l',Q_l,q_{l+1}}  a_{Q_l'} \prod_{\lambda=1}^l U_{q'_\lambda,r_\lambda}  \prod_{\lambda=1}^{l+1} U^{\dagger}_{r_\lambda,q_\lambda}  \, a_{Q_lq_{l+1}} \delta_{l,\mathrm{even}}
     + \sum_{R_l, r_{l+1}}\sum_{Q_l',Q_l,q_{l+1}}  \prod_{\lambda=1}^{l} U^{\dagger}_{r_\lambda,q_\lambda}  \, a_{Q_l} a_{Q_l',q'_{l+1}} \prod_{\lambda=1}^{l+1} U_{q'_\lambda,r_\lambda}   \delta_{l,\mathrm{odd}}\\
    & = \sum_{r_{l+1}}\sum_{Q_l',Q_l,q_{l+1}}  \delta_{Q_l',Q_l} a_{Q_l'}  U^{\dagger}_{r_{l+1},q_{l+1}} a_{Q_lq_{l+1}} \delta_{l,\mathrm{even}}
     + \sum_{r_{l+1}}\sum_{Q_l',Q_l,q_{l+1}} \delta_{Q_l',Q_l} a_{Q_l}   a_{Q'_l q'_{l+1}} U_{q'_{l+1},r_{l+1}} \delta_{l,\mathrm{odd}} \\
    & = \sum_{Q_l,q,r}\frac{1}{\sqrt{\tilde{z}}} e^{-2\pi \mathrm{i} \cdot q r} a_{Q_l} a_{Q_lq} \delta_{l,\mathrm{even}}
      + \sum_{Q_l,q,r}\frac{1}{\sqrt{\tilde{z}}} e^{+2\pi \mathrm{i} \cdot q r} a_{Q_l} a_{Q_lq} \delta_{l,\mathrm{even}}
     = \sqrt{\tilde{z}} \sum_{Q_l} a_{Q_l} a_{Q_l0},
\end{split}
\end{equation}
which again falls into an effective 1D chain geometry.
\end{widetext}

\section{The Bogoliubov transformation in the XY model}
\label{App:Bogoliubov}
  In this appendix, we provide the details of the Bogoliubov transformation for the XY model.
  We start from the 1D half-chain bosonic Hamiltonian, i.e., Eq.~(\ref{Eq:XYboson}) in the main text. By ignoring the ``impurity site" at the center, performing a Fourier transformation on an infinite chain (without changing the energy dispersion curve), we arrive at $H_{\rm{1D}} = \sum_{\kappa \ge 0} H_\kappa$, where
\begin{equation}
  H_\kappa = S[(z + \gamma_\kappa ) (b_\kappa^\dagger b_\kappa + b_{-\kappa}^\dagger b_{-\kappa}) +\gamma_\kappa (b_\kappa b_{-\kappa} + b_\kappa^\dagger b_{-\kappa}^\dagger)]
\end{equation}
with $\gamma_\kappa = \sqrt{z-1}\operatorname{cos}\kappa$.

The sub-Hamiltonian $H_\kappa$ can be rewritten as $H_\kappa = \frac{1}{2}B^\dagger \mathcal{H}_\kappa B$, where $B = (b_\kappa, b_{-\kappa}, b_\kappa^\dagger, b_{-\kappa}^\dagger)^T$, and
\begin{equation}
\label{Eq:XYHk}
  \mathcal{H}_\kappa =
  \begin{pmatrix}
  z+\gamma_\kappa  & & & \gamma_\kappa \\
    & z+\gamma_\kappa & \gamma_\kappa & \\
    & \gamma_\kappa & z+\gamma_\kappa & \\
  \gamma_\kappa & & & z+\gamma_\kappa \\
  \end{pmatrix},
\end{equation}
  where we have omitted the term $- (z+\gamma_\kappa)$ which is a constant after summing over $\kappa$.

In the Bogoliubov transformation, we find a matrix $\Theta$ so that (1) $B = \Theta\tilde{B}$, where $\tilde{B} = (\beta_\kappa, \beta_{-\kappa}, \beta_\kappa^\dagger, \beta_{-\kappa}^\dagger)^T$ still represents boson operators obeying the bosonic statistics, and (2) $H_\kappa = \frac{1}{2}\tilde{B}^\dagger \Theta^\dagger \mathcal{H}_\kappa \Theta \tilde{B} = \frac{1}{2}\tilde{B}^\dagger \mathbf{D} \tilde{B}$ is in a diagonal form.

To maintain the boson statistics in condition (1), we require $\Theta^\dagger \Lambda_z \Theta = \Lambda_z$, where
\begin{equation*}
 \Lambda_z = \begin{pmatrix}
  \mathbf{I}_{2\times2} & \\
    & -\mathbf{I}_{2\times2} \\
  \end{pmatrix}.
\end{equation*}
and $\mathbf{I}_{2\times2}$ is a $2\times 2$ identity matrix.

Combing together conditions (1) and (2), we have $(\Lambda_z \mathcal{H}_\kappa) \Theta = \Theta \Lambda_z \Theta^\dagger \mathcal{H}_\kappa \Theta = \Theta \Lambda_z \mathbf{D}$, which implies $\Theta$ and $\mathbf{D}$ can be found by diagonalizing $\Lambda_z\mathcal{H}_\kappa$.
After some calculations, and by observing that the Hamiltonian matrix in Eq.~(\ref{Eq:XYHk}) is block diagonal, we arrive at
\begin{equation*}
  \tilde{\epsilon}^2 - (z+\gamma_\kappa)^2 + \gamma_\kappa^2 = 0,
\end{equation*}
  where the positive root with $\kappa>0$ constitutes the magnon spectrum in Eq.~(\ref{Eq:BandXY}), i.e., $\epsilon(\kappa) = \tilde{\epsilon}(\kappa)$.
The corresponding Bogolon (annihilation) operator turns out to be
\begin{equation}
\begin{split}
  \beta_\kappa = \operatorname{cosh}(\theta_\kappa) b_\kappa - \operatorname{sinh}(\theta_\kappa) b_{-\kappa}^\dagger, \\
\end{split}
\label{Eq:Bogolon}
\end{equation}
with restriction $\kappa\in [0,\pi]$, where $\theta_\kappa$ can be determined from $\operatorname{tanh}(\theta_\kappa) = -\gamma_\kappa/(z+\gamma_\kappa)$.

\end{appendix}

\bibliography{BetheRefs}

\begin{thebibliography}{48}%
\makeatletter
\providecommand \@ifxundefined [1]{%
 \@ifx{#1\undefined}
}%
\providecommand \@ifnum [1]{%
 \ifnum #1\expandafter \@firstoftwo
 \else \expandafter \@secondoftwo
 \fi
}%
\providecommand \@ifx [1]{%
 \ifx #1\expandafter \@firstoftwo
 \else \expandafter \@secondoftwo
 \fi
}%
\providecommand \natexlab [1]{#1}%
\providecommand \enquote  [1]{``#1''}%
\providecommand \bibnamefont  [1]{#1}%
\providecommand \bibfnamefont [1]{#1}%
\providecommand \citenamefont [1]{#1}%
\providecommand \href@noop [0]{\@secondoftwo}%
\providecommand \href [0]{\begingroup \@sanitize@url \@href}%
\providecommand \@href[1]{\@@startlink{#1}\@@href}%
\providecommand \@@href[1]{\endgroup#1\@@endlink}%
\providecommand \@sanitize@url [0]{\catcode `\\12\catcode `\$12\catcode
  `\&12\catcode `\#12\catcode `\^12\catcode `\_12\catcode `\%12\relax}%
\providecommand \@@startlink[1]{}%
\providecommand \@@endlink[0]{}%
\providecommand \url  [0]{\begingroup\@sanitize@url \@url }%
\providecommand \@url [1]{\endgroup\@href {#1}{\urlprefix }}%
\providecommand \urlprefix  [0]{URL }%
\providecommand \Eprint [0]{\href }%
\providecommand \doibase [0]{http://dx.doi.org/}%
\providecommand \selectlanguage [0]{\@gobble}%
\providecommand \bibinfo  [0]{\@secondoftwo}%
\providecommand \bibfield  [0]{\@secondoftwo}%
\providecommand \translation [1]{[#1]}%
\providecommand \BibitemOpen [0]{}%
\providecommand \bibitemStop [0]{}%
\providecommand \bibitemNoStop [0]{.\EOS\space}%
\providecommand \EOS [0]{\spacefactor3000\relax}%
\providecommand \BibitemShut  [1]{\csname bibitem#1\endcsname}%
\let\auto@bib@innerbib\@empty
\bibitem [{\citenamefont {Bethe}(1935)}]{Bethe1935}%
  \BibitemOpen
  \bibfield  {author} {\bibinfo {author} {\bibfnamefont {H.~A.}\ \bibnamefont
  {Bethe}},\ }\bibfield  {title} {\enquote {\bibinfo {title} {Statistical
  theory of superlattices},}\ }\href {\doibase 10.1098/rspa.1935.0122}
  {\bibfield  {journal} {\bibinfo  {journal} {Proc. R. Soc. A}\ }\textbf
  {\bibinfo {volume} {150}},\ \bibinfo {pages} {552} (\bibinfo {year}
  {1935})}\BibitemShut {NoStop}%
\bibitem [{\citenamefont {Pathria}\ and\ \citenamefont
  {Beale}(2011)}]{Pathria2011}%
  \BibitemOpen
  \bibfield  {author} {\bibinfo {author} {\bibfnamefont {R.~K.}\ \bibnamefont
  {Pathria}}\ and\ \bibinfo {author} {\bibfnamefont {P.~D.}\ \bibnamefont
  {Beale}},\ }\href@noop {} {\emph {\bibinfo {title} {Statistical
  {Mechanics}}}}\ (\bibinfo  {publisher} {{Academic Press; 3 edition}},\
  \bibinfo {year} {2011})\BibitemShut {NoStop}%
\bibitem [{\citenamefont {Ostilli}(2012)}]{Ostilli2012}%
  \BibitemOpen
  \bibfield  {author} {\bibinfo {author} {\bibfnamefont {M.}~\bibnamefont
  {Ostilli}},\ }\bibfield  {title} {\enquote {\bibinfo {title} {Cayley trees
  and {Bethe} lattices: A concise analysis for mathematicians and
  physicists},}\ }\href {\doibase 10.1016/j.physa.2012.01.038} {\bibfield
  {journal} {\bibinfo  {journal} {Physica A}\ }\textbf {\bibinfo {volume}
  {391}},\ \bibinfo {pages} {3417} (\bibinfo {year} {2012})}\BibitemShut
  {NoStop}%
\bibitem [{\citenamefont {Eckstein}\ \emph {et~al.}(2005)\citenamefont
  {Eckstein}, \citenamefont {Kollar}, \citenamefont {Byczuk},\ and\
  \citenamefont {Vollhardt}}]{Eckstein2005}%
  \BibitemOpen
  \bibfield  {author} {\bibinfo {author} {\bibfnamefont {M.}~\bibnamefont
  {Eckstein}}, \bibinfo {author} {\bibfnamefont {M.}~\bibnamefont {Kollar}},
  \bibinfo {author} {\bibfnamefont {K.}~\bibnamefont {Byczuk}}, \ and\ \bibinfo
  {author} {\bibfnamefont {D.}~\bibnamefont {Vollhardt}},\ }\bibfield  {title}
  {\enquote {\bibinfo {title} {Hopping on the {Bethe} lattice: Exact results
  for densities of states and dynamical mean-field theory},}\ }\href {\doibase
  10.1103/PhysRevB.71.235119} {\bibfield  {journal} {\bibinfo  {journal} {Phys.
  Rev. B}\ }\textbf {\bibinfo {volume} {71}},\ \bibinfo {pages} {235119}
  (\bibinfo {year} {2005})}\BibitemShut {NoStop}%
\bibitem [{\citenamefont {Petrova}\ and\ \citenamefont
  {Moessner}(2016)}]{Olga2016}%
  \BibitemOpen
  \bibfield  {author} {\bibinfo {author} {\bibfnamefont {O.}~\bibnamefont
  {Petrova}}\ and\ \bibinfo {author} {\bibfnamefont {R.}~\bibnamefont
  {Moessner}},\ }\bibfield  {title} {\enquote {\bibinfo {title} {Coulomb
  potential {$V(r)=1/r$} problem on the bethe lattice},}\ }\href {\doibase
  10.1103/PhysRevE.93.012115} {\bibfield  {journal} {\bibinfo  {journal} {Phys.
  Rev. E}\ }\textbf {\bibinfo {volume} {93}},\ \bibinfo {pages} {012115}
  (\bibinfo {year} {2016})}\BibitemShut {NoStop}%
\bibitem [{\citenamefont {Semerjian}\ \emph {et~al.}(2009)\citenamefont
  {Semerjian}, \citenamefont {Tarzia},\ and\ \citenamefont
  {Zamponi}}]{Semerjian2009}%
  \BibitemOpen
  \bibfield  {author} {\bibinfo {author} {\bibfnamefont {G.}~\bibnamefont
  {Semerjian}}, \bibinfo {author} {\bibfnamefont {M.}~\bibnamefont {Tarzia}}, \
  and\ \bibinfo {author} {\bibfnamefont {F.}~\bibnamefont {Zamponi}},\
  }\bibfield  {title} {\enquote {\bibinfo {title} {Exact solution of the
  {Bose-Hubbard} model on the {Bethe} lattice},}\ }\href {\doibase
  10.1103/PhysRevB.80.014524} {\bibfield  {journal} {\bibinfo  {journal} {Phys.
  Rev. B}\ }\textbf {\bibinfo {volume} {80}},\ \bibinfo {pages} {014524}
  (\bibinfo {year} {2009})}\BibitemShut {NoStop}%
\bibitem [{\citenamefont {Jiang}\ \emph {et~al.}(2008)\citenamefont {Jiang},
  \citenamefont {Weng},\ and\ \citenamefont {Xiang}}]{Jiang2008}%
  \BibitemOpen
  \bibfield  {author} {\bibinfo {author} {\bibfnamefont {H.~C.}\ \bibnamefont
  {Jiang}}, \bibinfo {author} {\bibfnamefont {Z.~Y.}\ \bibnamefont {Weng}}, \
  and\ \bibinfo {author} {\bibfnamefont {T.}~\bibnamefont {Xiang}},\ }\bibfield
   {title} {\enquote {\bibinfo {title} {Accurate determination of tensor
  network state of quantum lattice models in two dimensions},}\ }\href
  {\doibase 10.1103/PhysRevLett.101.090603} {\bibfield  {journal} {\bibinfo
  {journal} {Phys. Rev. Lett.}\ }\textbf {\bibinfo {volume} {101}},\ \bibinfo
  {pages} {090603} (\bibinfo {year} {2008})}\BibitemShut {NoStop}%
\bibitem [{\citenamefont {Verstraete}\ \emph {et~al.}(2008)\citenamefont
  {Verstraete}, \citenamefont {Murg},\ and\ \citenamefont
  {Cirac}}]{Verstraete2008}%
  \BibitemOpen
  \bibfield  {author} {\bibinfo {author} {\bibfnamefont {F.}~\bibnamefont
  {Verstraete}}, \bibinfo {author} {\bibfnamefont {V.}~\bibnamefont {Murg}}, \
  and\ \bibinfo {author} {\bibfnamefont {J.~I.}\ \bibnamefont {Cirac}},\
  }\bibfield  {title} {\enquote {\bibinfo {title} {Matrix product states,
  projected entangled pair states, and variational renormalization group
  methods for quantum spin systems},}\ }\href {\doibase
  10.1080/14789940801912366} {\bibfield  {journal} {\bibinfo  {journal} {Adv.
  Phys.}\ }\textbf {\bibinfo {volume} {57}},\ \bibinfo {pages} {143} (\bibinfo
  {year} {2008})}\BibitemShut {NoStop}%
\bibitem [{\citenamefont {Or\'us}(2014)}]{Orus2014}%
  \BibitemOpen
  \bibfield  {author} {\bibinfo {author} {\bibfnamefont {R.}~\bibnamefont
  {Or\'us}},\ }\bibfield  {title} {\enquote {\bibinfo {title} {A practical
  introduction to tensor networks: Matrix product states and projected
  entangled pair states},}\ }\href {\doibase
  https://doi.org/10.1016/j.aop.2014.06.013} {\bibfield  {journal} {\bibinfo
  {journal} {Ann. Phys.}\ }\textbf {\bibinfo {volume} {349}},\ \bibinfo {pages}
  {117} (\bibinfo {year} {2014})}\BibitemShut {NoStop}%
\bibitem [{\citenamefont {Li}\ \emph {et~al.}(2012)\citenamefont {Li},
  \citenamefont {von Delft},\ and\ \citenamefont {Xiang}}]{Li2012}%
  \BibitemOpen
  \bibfield  {author} {\bibinfo {author} {\bibfnamefont {W.}~\bibnamefont
  {Li}}, \bibinfo {author} {\bibfnamefont {J.}~\bibnamefont {von Delft}}, \
  and\ \bibinfo {author} {\bibfnamefont {T.}~\bibnamefont {Xiang}},\ }\bibfield
   {title} {\enquote {\bibinfo {title} {Efficient simulation of infinite tree
  tensor network states on the {Bethe} lattice},}\ }\href {\doibase
  10.1103/PhysRevB.86.195137} {\bibfield  {journal} {\bibinfo  {journal} {Phys.
  Rev. B}\ }\textbf {\bibinfo {volume} {86}},\ \bibinfo {pages} {195137}
  (\bibinfo {year} {2012})}\BibitemShut {NoStop}%
\bibitem [{\citenamefont {Li}\ \emph {et~al.}(2010)\citenamefont {Li},
  \citenamefont {Gong}, \citenamefont {Zhao},\ and\ \citenamefont
  {Su}}]{Li2010}%
  \BibitemOpen
  \bibfield  {author} {\bibinfo {author} {\bibfnamefont {W.}~\bibnamefont
  {Li}}, \bibinfo {author} {\bibfnamefont {S.-S.}\ \bibnamefont {Gong}},
  \bibinfo {author} {\bibfnamefont {Y.}~\bibnamefont {Zhao}}, \ and\ \bibinfo
  {author} {\bibfnamefont {G.}~\bibnamefont {Su}},\ }\bibfield  {title}
  {\enquote {\bibinfo {title} {Quantum phase transition, {$O(3)$} universality
  class, and phase diagram of the spin-$\frac{1}{2}$ {Heisenberg}
  antiferromagnet on a distorted honeycomb lattice: A tensor
  renormalization-group study},}\ }\href {\doibase 10.1103/PhysRevB.81.184427}
  {\bibfield  {journal} {\bibinfo  {journal} {Phys. Rev. B}\ }\textbf {\bibinfo
  {volume} {81}},\ \bibinfo {pages} {184427} (\bibinfo {year}
  {2010})}\BibitemShut {NoStop}%
\bibitem [{\citenamefont {Liu}\ \emph {et~al.}(2014)\citenamefont {Liu},
  \citenamefont {Ran}, \citenamefont {Li}, \citenamefont {Yan}, \citenamefont
  {Zhao},\ and\ \citenamefont {Su}}]{Liu2014}%
  \BibitemOpen
  \bibfield  {author} {\bibinfo {author} {\bibfnamefont {T.}~\bibnamefont
  {Liu}}, \bibinfo {author} {\bibfnamefont {S.-J.}\ \bibnamefont {Ran}},
  \bibinfo {author} {\bibfnamefont {W.}~\bibnamefont {Li}}, \bibinfo {author}
  {\bibfnamefont {X.}~\bibnamefont {Yan}}, \bibinfo {author} {\bibfnamefont
  {Y.}~\bibnamefont {Zhao}}, \ and\ \bibinfo {author} {\bibfnamefont
  {G.}~\bibnamefont {Su}},\ }\bibfield  {title} {\enquote {\bibinfo {title}
  {Featureless quantum spin liquid, $\frac{1}{3}$-magnetization plateau state,
  and exotic thermodynamic properties of the spin-$\frac{1}{2}$ frustrated
  {Heisenberg} antiferromagnet on an infinite husimi lattice},}\ }\href
  {\doibase 10.1103/PhysRevB.89.054426} {\bibfield  {journal} {\bibinfo
  {journal} {Phys. Rev. B}\ }\textbf {\bibinfo {volume} {89}},\ \bibinfo
  {pages} {054426} (\bibinfo {year} {2014})}\BibitemShut {NoStop}%
\bibitem [{\citenamefont {Liu}\ \emph {et~al.}(2015)\citenamefont {Liu},
  \citenamefont {Li}, \citenamefont {Weichselbaum}, \citenamefont {von Delft},\
  and\ \citenamefont {Su}}]{Liu2015}%
  \BibitemOpen
  \bibfield  {author} {\bibinfo {author} {\bibfnamefont {T.}~\bibnamefont
  {Liu}}, \bibinfo {author} {\bibfnamefont {W.}~\bibnamefont {Li}}, \bibinfo
  {author} {\bibfnamefont {A.}~\bibnamefont {Weichselbaum}}, \bibinfo {author}
  {\bibfnamefont {J.}~\bibnamefont {von Delft}}, \ and\ \bibinfo {author}
  {\bibfnamefont {G.}~\bibnamefont {Su}},\ }\bibfield  {title} {\enquote
  {\bibinfo {title} {Simplex valence-bond crystal in the spin-1 kagome
  {Heisenberg} antiferromagnet},}\ }\href {\doibase 10.1103/PhysRevB.91.060403}
  {\bibfield  {journal} {\bibinfo  {journal} {Phys. Rev. B}\ }\textbf {\bibinfo
  {volume} {91}},\ \bibinfo {pages} {060403} (\bibinfo {year}
  {2015})}\BibitemShut {NoStop}%
\bibitem [{\citenamefont {Liu}\ \emph {et~al.}(2016)\citenamefont {Liu},
  \citenamefont {Li},\ and\ \citenamefont {Su}}]{Liu2016}%
  \BibitemOpen
  \bibfield  {author} {\bibinfo {author} {\bibfnamefont {T.}~\bibnamefont
  {Liu}}, \bibinfo {author} {\bibfnamefont {W.}~\bibnamefont {Li}}, \ and\
  \bibinfo {author} {\bibfnamefont {G.}~\bibnamefont {Su}},\ }\bibfield
  {title} {\enquote {\bibinfo {title} {Spin-ordered ground state and
  thermodynamic behaviors of the spin-$\frac{3}{2}$ kagome {Heisenberg}
  antiferromagnet},}\ }\href {\doibase 10.1103/PhysRevE.94.032114} {\bibfield
  {journal} {\bibinfo  {journal} {Phys. Rev. E}\ }\textbf {\bibinfo {volume}
  {94}},\ \bibinfo {pages} {032114} (\bibinfo {year} {2016})}\BibitemShut
  {NoStop}%
\bibitem [{\citenamefont {Otsuka}(1996)}]{Otsuka1996}%
  \BibitemOpen
  \bibfield  {author} {\bibinfo {author} {\bibfnamefont {H.}~\bibnamefont
  {Otsuka}},\ }\bibfield  {title} {\enquote {\bibinfo {title} {Density-matrix
  renormalization-group study of the spin-1/2 $\mathrm{XXZ}$ antiferromagnet on
  the {Bethe} lattice},}\ }\href {\doibase 10.1103/PhysRevB.53.14004}
  {\bibfield  {journal} {\bibinfo  {journal} {Phys. Rev. B}\ }\textbf {\bibinfo
  {volume} {53}},\ \bibinfo {pages} {14004} (\bibinfo {year}
  {1996})}\BibitemShut {NoStop}%
\bibitem [{\citenamefont {Friedman}(1997)}]{Friedman1997}%
  \BibitemOpen
  \bibfield  {author} {\bibinfo {author} {\bibfnamefont {B.}~\bibnamefont
  {Friedman}},\ }\bibfield  {title} {\enquote {\bibinfo {title} {A density
  matrix renormalization group approach to interacting quantum systems on
  {Cayley} trees},}\ }\href {\doibase 10.1088/0953-8984/9/42/016} {\bibfield
  {journal} {\bibinfo  {journal} {J. Phys.: Condens. Matter}\ }\textbf
  {\bibinfo {volume} {9}},\ \bibinfo {pages} {9021} (\bibinfo {year}
  {1997})}\BibitemShut {NoStop}%
\bibitem [{\citenamefont {Kumar}\ \emph {et~al.}(2012)\citenamefont {Kumar},
  \citenamefont {Ramasesha},\ and\ \citenamefont {Soos}}]{Kumar2012}%
  \BibitemOpen
  \bibfield  {author} {\bibinfo {author} {\bibfnamefont {M.}~\bibnamefont
  {Kumar}}, \bibinfo {author} {\bibfnamefont {S.}~\bibnamefont {Ramasesha}}, \
  and\ \bibinfo {author} {\bibfnamefont {Z.~G.}\ \bibnamefont {Soos}},\
  }\bibfield  {title} {\enquote {\bibinfo {title} {Density matrix
  renormalization group algorithm for bethe lattices of spin-$\frac{1}{2}$ or
  spin-1 sites with {Heisenberg} antiferromagnetic exchange},}\ }\href
  {\doibase 10.1103/PhysRevB.85.134415} {\bibfield  {journal} {\bibinfo
  {journal} {Phys. Rev. B}\ }\textbf {\bibinfo {volume} {85}},\ \bibinfo
  {pages} {134415} (\bibinfo {year} {2012})}\BibitemShut {NoStop}%
\bibitem [{\citenamefont {Changlani}\ \emph
  {et~al.}(2013{\natexlab{a}})\citenamefont {Changlani}, \citenamefont {Ghosh},
  \citenamefont {Henley},\ and\ \citenamefont {L\"auchli}}]{Changlani2013}%
  \BibitemOpen
  \bibfield  {author} {\bibinfo {author} {\bibfnamefont {H.~J.}\ \bibnamefont
  {Changlani}}, \bibinfo {author} {\bibfnamefont {S.}~\bibnamefont {Ghosh}},
  \bibinfo {author} {\bibfnamefont {C.~L.}\ \bibnamefont {Henley}}, \ and\
  \bibinfo {author} {\bibfnamefont {A.~M.}\ \bibnamefont {L\"auchli}},\
  }\bibfield  {title} {\enquote {\bibinfo {title} {{Heisenberg} antiferromagnet
  on cayley trees: Low-energy spectrum and even/odd site imbalance},}\ }\href
  {\doibase 10.1103/PhysRevB.87.085107} {\bibfield  {journal} {\bibinfo
  {journal} {Phys. Rev. B}\ }\textbf {\bibinfo {volume} {87}},\ \bibinfo
  {pages} {085107} (\bibinfo {year} {2013}{\natexlab{a}})}\BibitemShut
  {NoStop}%
\bibitem [{\citenamefont {Changlani}\ \emph
  {et~al.}(2013{\natexlab{b}})\citenamefont {Changlani}, \citenamefont {Ghosh},
  \citenamefont {Pujari},\ and\ \citenamefont {Henley}}]{Changlani2013PRL}%
  \BibitemOpen
  \bibfield  {author} {\bibinfo {author} {\bibfnamefont {H.~J.}\ \bibnamefont
  {Changlani}}, \bibinfo {author} {\bibfnamefont {S.}~\bibnamefont {Ghosh}},
  \bibinfo {author} {\bibfnamefont {S.}~\bibnamefont {Pujari}}, \ and\ \bibinfo
  {author} {\bibfnamefont {C.~L.}\ \bibnamefont {Henley}},\ }\bibfield  {title}
  {\enquote {\bibinfo {title} {Emergent spin excitations in a {Bethe} lattice
  at percolation},}\ }\href {\doibase 10.1103/PhysRevLett.111.157201}
  {\bibfield  {journal} {\bibinfo  {journal} {Phys. Rev. Lett.}\ }\textbf
  {\bibinfo {volume} {111}},\ \bibinfo {pages} {157201} (\bibinfo {year}
  {2013}{\natexlab{b}})}\BibitemShut {NoStop}%
\bibitem [{\citenamefont {Brinkman}\ and\ \citenamefont
  {Rice}(1970)}]{Brinkman1970}%
  \BibitemOpen
  \bibfield  {author} {\bibinfo {author} {\bibfnamefont {W.~F.}\ \bibnamefont
  {Brinkman}}\ and\ \bibinfo {author} {\bibfnamefont {T.~M.}\ \bibnamefont
  {Rice}},\ }\bibfield  {title} {\enquote {\bibinfo {title} {Single-particle
  excitations in magnetic insulators},}\ }\href {\doibase
  10.1103/PhysRevB.2.1324} {\bibfield  {journal} {\bibinfo  {journal} {Phys.
  Rev. B}\ }\textbf {\bibinfo {volume} {2}},\ \bibinfo {pages} {1324} (\bibinfo
  {year} {1970})}\BibitemShut {NoStop}%
\bibitem [{\citenamefont {Kittler}\ and\ \citenamefont
  {Falicov}(1976)}]{Kittler1976}%
  \BibitemOpen
  \bibfield  {author} {\bibinfo {author} {\bibfnamefont {R.~C.}\ \bibnamefont
  {Kittler}}\ and\ \bibinfo {author} {\bibfnamefont {L.~M.}\ \bibnamefont
  {Falicov}},\ }\bibfield  {title} {\enquote {\bibinfo {title} {Electronic
  structure of disordered binary alloys},}\ }\href {\doibase
  10.1088/0022-3719/9/23/010} {\bibfield  {journal} {\bibinfo  {journal} {J.
  Phys. C: Solid State Phys.}\ }\textbf {\bibinfo {volume} {9}},\ \bibinfo
  {pages} {4259} (\bibinfo {year} {1976})}\BibitemShut {NoStop}%
\bibitem [{\citenamefont {Brouers}\ and\ \citenamefont
  {Marconi}(1982)}]{Brouers1982}%
  \BibitemOpen
  \bibfield  {author} {\bibinfo {author} {\bibfnamefont {F.}~\bibnamefont
  {Brouers}}\ and\ \bibinfo {author} {\bibfnamefont {U.~M.~B.}\ \bibnamefont
  {Marconi}},\ }\bibfield  {title} {\enquote {\bibinfo {title} {On the
  antiferromagnetic phase in the {Hubbard} model},}\ }\href@noop {} {\bibfield
  {journal} {\bibinfo  {journal} {J. Phys. C: Solid State Phys.}\ }\textbf
  {\bibinfo {volume} {15}},\ \bibinfo {pages} {L925} (\bibinfo {year}
  {1982})}\BibitemShut {NoStop}%
\bibitem [{\citenamefont {Al~Hajj}\ \emph {et~al.}(2004)\citenamefont
  {Al~Hajj}, \citenamefont {Guih\'ery}, \citenamefont {Malrieu},\ and\
  \citenamefont {Wind}}]{Wind2004}%
  \BibitemOpen
  \bibfield  {author} {\bibinfo {author} {\bibfnamefont {M.}~\bibnamefont
  {Al~Hajj}}, \bibinfo {author} {\bibfnamefont {N.}~\bibnamefont {Guih\'ery}},
  \bibinfo {author} {\bibfnamefont {J.-P.}\ \bibnamefont {Malrieu}}, \ and\
  \bibinfo {author} {\bibfnamefont {P.}~\bibnamefont {Wind}},\ }\bibfield
  {title} {\enquote {\bibinfo {title} {Theoretical studies of the phase
  transition in the anisotropic two-dimensional square spin lattice},}\ }\href
  {\doibase 10.1103/PhysRevB.70.094415} {\bibfield  {journal} {\bibinfo
  {journal} {Phys. Rev. B}\ }\textbf {\bibinfo {volume} {70}},\ \bibinfo
  {pages} {094415} (\bibinfo {year} {2004})}\BibitemShut {NoStop}%
\bibitem [{\citenamefont {White}(1992)}]{PhysRevLett.69.2863}%
  \BibitemOpen
  \bibfield  {author} {\bibinfo {author} {\bibfnamefont {S.~R.}\ \bibnamefont
  {White}},\ }\bibfield  {title} {\enquote {\bibinfo {title} {Density matrix
  formulation for quantum renormalization groups},}\ }\href {\doibase
  10.1103/PhysRevLett.69.2863} {\bibfield  {journal} {\bibinfo  {journal}
  {Phys. Rev. Lett.}\ }\textbf {\bibinfo {volume} {69}},\ \bibinfo {pages}
  {2863} (\bibinfo {year} {1992})}\BibitemShut {NoStop}%
\bibitem [{\citenamefont {Vidal}(2007)}]{Vidal2007}%
  \BibitemOpen
  \bibfield  {author} {\bibinfo {author} {\bibfnamefont {G.}~\bibnamefont
  {Vidal}},\ }\bibfield  {title} {\enquote {\bibinfo {title} {Classical
  simulation of infinite-size quantum lattice systems in one spatial
  dimension},}\ }\href {\doibase 10.1103/PhysRevLett.98.070201} {\bibfield
  {journal} {\bibinfo  {journal} {Phys. Rev. Lett.}\ }\textbf {\bibinfo
  {volume} {98}},\ \bibinfo {pages} {070201} (\bibinfo {year}
  {2007})}\BibitemShut {NoStop}%
\bibitem [{\citenamefont {Or\'us}\ and\ \citenamefont
  {Vidal}(2008)}]{OrusVidal2008}%
  \BibitemOpen
  \bibfield  {author} {\bibinfo {author} {\bibfnamefont {R.}~\bibnamefont
  {Or\'us}}\ and\ \bibinfo {author} {\bibfnamefont {G.}~\bibnamefont {Vidal}},\
  }\bibfield  {title} {\enquote {\bibinfo {title} {Infinite time-evolving block
  decimation algorithm beyond unitary evolution},}\ }\href {\doibase
  10.1103/PhysRevB.78.155117} {\bibfield  {journal} {\bibinfo  {journal} {Phys.
  Rev. B}\ }\textbf {\bibinfo {volume} {78}},\ \bibinfo {pages} {155117}
  (\bibinfo {year} {2008})}\BibitemShut {NoStop}%
\bibitem [{\citenamefont {Li}\ \emph {et~al.}(2011)\citenamefont {Li},
  \citenamefont {Ran}, \citenamefont {Gong}, \citenamefont {Zhao},
  \citenamefont {Xi}, \citenamefont {Ye},\ and\ \citenamefont
  {Su}}]{Li.w+:2011:LTRG}%
  \BibitemOpen
  \bibfield  {author} {\bibinfo {author} {\bibfnamefont {W.}~\bibnamefont
  {Li}}, \bibinfo {author} {\bibfnamefont {S.-J.}\ \bibnamefont {Ran}},
  \bibinfo {author} {\bibfnamefont {S.-S.}\ \bibnamefont {Gong}}, \bibinfo
  {author} {\bibfnamefont {Y.}~\bibnamefont {Zhao}}, \bibinfo {author}
  {\bibfnamefont {B.}~\bibnamefont {Xi}}, \bibinfo {author} {\bibfnamefont
  {F.}~\bibnamefont {Ye}}, \ and\ \bibinfo {author} {\bibfnamefont
  {G.}~\bibnamefont {Su}},\ }\bibfield  {title} {\enquote {\bibinfo {title}
  {Linearized tensor renormalization group algorithm for the calculation of
  thermodynamic properties of quantum lattice models},}\ }\href {\doibase
  10.1103/PhysRevLett.106.127202} {\bibfield  {journal} {\bibinfo  {journal}
  {Phys. Rev. Lett.}\ }\textbf {\bibinfo {volume} {106}},\ \bibinfo {pages}
  {127202} (\bibinfo {year} {2011})}\BibitemShut {NoStop}%
\bibitem [{\citenamefont {Dong}\ \emph {et~al.}(2017)\citenamefont {Dong},
  \citenamefont {Chen}, \citenamefont {Liu},\ and\ \citenamefont
  {Li}}]{Dong.y+:2017:BiLTRG}%
  \BibitemOpen
  \bibfield  {author} {\bibinfo {author} {\bibfnamefont {Y.-L.}\ \bibnamefont
  {Dong}}, \bibinfo {author} {\bibfnamefont {L.}~\bibnamefont {Chen}}, \bibinfo
  {author} {\bibfnamefont {Y.-J.}\ \bibnamefont {Liu}}, \ and\ \bibinfo
  {author} {\bibfnamefont {W.}~\bibnamefont {Li}},\ }\bibfield  {title}
  {\enquote {\bibinfo {title} {Bilayer linearized tensor renormalization group
  approach for thermal tensor networks},}\ }\href {\doibase
  10.1103/PhysRevB.95.144428} {\bibfield  {journal} {\bibinfo  {journal} {Phys.
  Rev. B}\ }\textbf {\bibinfo {volume} {95}},\ \bibinfo {pages} {144428}
  (\bibinfo {year} {2017})}\BibitemShut {NoStop}%
\bibitem [{\citenamefont {Ran}\ \emph {et~al.}(2012)\citenamefont {Ran},
  \citenamefont {Li}, \citenamefont {Xi}, \citenamefont {Zhang},\ and\
  \citenamefont {Su}}]{Ran.s+:2012:Super-orthogonalization}%
  \BibitemOpen
  \bibfield  {author} {\bibinfo {author} {\bibfnamefont {S.-J.}\ \bibnamefont
  {Ran}}, \bibinfo {author} {\bibfnamefont {W.}~\bibnamefont {Li}}, \bibinfo
  {author} {\bibfnamefont {B.}~\bibnamefont {Xi}}, \bibinfo {author}
  {\bibfnamefont {Z.}~\bibnamefont {Zhang}}, \ and\ \bibinfo {author}
  {\bibfnamefont {G.}~\bibnamefont {Su}},\ }\bibfield  {title} {\enquote
  {\bibinfo {title} {Optimized decimation of tensor networks with
  super-orthogonalization for two-dimensional quantum lattice models},}\ }\href
  {\doibase 10.1103/PhysRevB.86.134429} {\bibfield  {journal} {\bibinfo
  {journal} {Phys. Rev. B}\ }\textbf {\bibinfo {volume} {86}},\ \bibinfo
  {pages} {134429} (\bibinfo {year} {2012})}\BibitemShut {NoStop}%
\bibitem [{\citenamefont {Chen}\ \emph {et~al.}(2017)\citenamefont {Chen},
  \citenamefont {Liu}, \citenamefont {Chen},\ and\ \citenamefont
  {Li}}]{Chen.b+:2017:SETTN}%
  \BibitemOpen
  \bibfield  {author} {\bibinfo {author} {\bibfnamefont {B.-B.}\ \bibnamefont
  {Chen}}, \bibinfo {author} {\bibfnamefont {Y.-J.}\ \bibnamefont {Liu}},
  \bibinfo {author} {\bibfnamefont {Z.}~\bibnamefont {Chen}}, \ and\ \bibinfo
  {author} {\bibfnamefont {W.}~\bibnamefont {Li}},\ }\bibfield  {title}
  {\enquote {\bibinfo {title} {Series-expansion thermal tensor network approach
  for quantum lattice models},}\ }\href {\doibase 10.1103/PhysRevB.95.161104}
  {\bibfield  {journal} {\bibinfo  {journal} {Phys. Rev. B}\ }\textbf {\bibinfo
  {volume} {95}},\ \bibinfo {pages} {161104} (\bibinfo {year}
  {2017})}\BibitemShut {NoStop}%
\bibitem [{\citenamefont {Chen}\ \emph {et~al.}(2018)\citenamefont {Chen},
  \citenamefont {Chen}, \citenamefont {Chen}, \citenamefont {Li},\ and\
  \citenamefont {Weichselbaum}}]{Chen2018}%
  \BibitemOpen
  \bibfield  {author} {\bibinfo {author} {\bibfnamefont {B.-B.}\ \bibnamefont
  {Chen}}, \bibinfo {author} {\bibfnamefont {L.}~\bibnamefont {Chen}}, \bibinfo
  {author} {\bibfnamefont {Z.}~\bibnamefont {Chen}}, \bibinfo {author}
  {\bibfnamefont {W.}~\bibnamefont {Li}}, \ and\ \bibinfo {author}
  {\bibfnamefont {A.}~\bibnamefont {Weichselbaum}},\ }\bibfield  {title}
  {\enquote {\bibinfo {title} {Exponential thermal tensor network approach for
  quantum lattice models},}\ }\href {\doibase 10.1103/PhysRevX.8.031082}
  {\bibfield  {journal} {\bibinfo  {journal} {Phys. Rev. X}\ }\textbf {\bibinfo
  {volume} {8}},\ \bibinfo {pages} {031082} (\bibinfo {year}
  {2018})}\BibitemShut {NoStop}%
\bibitem [{\citenamefont {Czarnik}\ \emph {et~al.}(2012)\citenamefont
  {Czarnik}, \citenamefont {Cincio},\ and\ \citenamefont
  {Dziarmaga}}]{Czarnik.p+:2012:PEPS}%
  \BibitemOpen
  \bibfield  {author} {\bibinfo {author} {\bibfnamefont {P.}~\bibnamefont
  {Czarnik}}, \bibinfo {author} {\bibfnamefont {L.}~\bibnamefont {Cincio}}, \
  and\ \bibinfo {author} {\bibfnamefont {J.}~\bibnamefont {Dziarmaga}},\
  }\bibfield  {title} {\enquote {\bibinfo {title} {Projected entangled pair
  states at finite temperature: Imaginary time evolution with ancillas},}\
  }\href {\doibase 10.1103/PhysRevB.86.245101} {\bibfield  {journal} {\bibinfo
  {journal} {Phys. Rev. B}\ }\textbf {\bibinfo {volume} {86}},\ \bibinfo
  {pages} {245101} (\bibinfo {year} {2012})}\BibitemShut {NoStop}%
\bibitem [{\citenamefont {Czarnik}\ and\ \citenamefont
  {Dziarmaga}(2015)}]{Czarnik.p+:2015:PEPS}%
  \BibitemOpen
  \bibfield  {author} {\bibinfo {author} {\bibfnamefont {P.}~\bibnamefont
  {Czarnik}}\ and\ \bibinfo {author} {\bibfnamefont {J.}~\bibnamefont
  {Dziarmaga}},\ }\bibfield  {title} {\enquote {\bibinfo {title} {Variational
  approach to projected entangled pair states at finite temperature},}\ }\href
  {\doibase 10.1103/PhysRevB.92.035152} {\bibfield  {journal} {\bibinfo
  {journal} {Phys. Rev. B}\ }\textbf {\bibinfo {volume} {92}},\ \bibinfo
  {pages} {035152} (\bibinfo {year} {2015})}\BibitemShut {NoStop}%
\bibitem [{\citenamefont {Czarnik}\ \emph {et~al.}(2016)\citenamefont
  {Czarnik}, \citenamefont {Rams},\ and\ \citenamefont
  {Dziarmaga}}]{Czarnik.p+:2016:TNR}%
  \BibitemOpen
  \bibfield  {author} {\bibinfo {author} {\bibfnamefont {P.}~\bibnamefont
  {Czarnik}}, \bibinfo {author} {\bibfnamefont {M.~M.}\ \bibnamefont {Rams}}, \
  and\ \bibinfo {author} {\bibfnamefont {J.}~\bibnamefont {Dziarmaga}},\
  }\bibfield  {title} {\enquote {\bibinfo {title} {Variational tensor network
  renormalization in imaginary time: Benchmark results in the {Hubbard} model
  at finite temperature},}\ }\href {\doibase 10.1103/PhysRevB.94.235142}
  {\bibfield  {journal} {\bibinfo  {journal} {Phys. Rev. B}\ }\textbf {\bibinfo
  {volume} {94}},\ \bibinfo {pages} {235142} (\bibinfo {year}
  {2016})}\BibitemShut {NoStop}%
\bibitem [{\citenamefont {Czarnik}\ \emph {et~al.}(2017)\citenamefont
  {Czarnik}, \citenamefont {Dziarmaga},\ and\ \citenamefont
  {Ole\ifmmode~\acute{s}\else \'{s}\fi{}}}]{Czarnik.p+:2017:Sign}%
  \BibitemOpen
  \bibfield  {author} {\bibinfo {author} {\bibfnamefont {P.}~\bibnamefont
  {Czarnik}}, \bibinfo {author} {\bibfnamefont {J.}~\bibnamefont {Dziarmaga}},
  \ and\ \bibinfo {author} {\bibfnamefont {A.~M.}\ \bibnamefont
  {Ole\ifmmode~\acute{s}\else \'{s}\fi{}}},\ }\bibfield  {title} {\enquote
  {\bibinfo {title} {Overcoming the sign problem at finite temperature: Quantum
  tensor network for the orbital ${e}_{g}$ model on an infinite square
  lattice},}\ }\href {\doibase 10.1103/PhysRevB.96.014420} {\bibfield
  {journal} {\bibinfo  {journal} {Phys. Rev. B}\ }\textbf {\bibinfo {volume}
  {96}},\ \bibinfo {pages} {014420} (\bibinfo {year} {2017})}\BibitemShut
  {NoStop}%
\bibitem [{\citenamefont {Kshetrimayum}\ \emph {et~al.}(2019)\citenamefont
  {Kshetrimayum}, \citenamefont {Rizzi}, \citenamefont {Eisert},\ and\
  \citenamefont {Or\'us}}]{Orus2018}%
  \BibitemOpen
  \bibfield  {author} {\bibinfo {author} {\bibfnamefont {A.}~\bibnamefont
  {Kshetrimayum}}, \bibinfo {author} {\bibfnamefont {M.}~\bibnamefont {Rizzi}},
  \bibinfo {author} {\bibfnamefont {J.}~\bibnamefont {Eisert}}, \ and\ \bibinfo
  {author} {\bibfnamefont {R.}~\bibnamefont {Or\'us}},\ }\bibfield  {title}
  {\enquote {\bibinfo {title} {Tensor network annealing algorithm for
  two-dimensional thermal states},}\ }\href {\doibase
  10.1103/PhysRevLett.122.070502} {\bibfield  {journal} {\bibinfo  {journal}
  {Phys. Rev. Lett.}\ }\textbf {\bibinfo {volume} {122}},\ \bibinfo {pages}
  {070502} (\bibinfo {year} {2019})}\BibitemShut {NoStop}%
\bibitem [{\citenamefont {Czarnik}\ \emph {et~al.}(2019)\citenamefont
  {Czarnik}, \citenamefont {Dziarmaga},\ and\ \citenamefont
  {Corboz}}]{Czarnik2019a}%
  \BibitemOpen
  \bibfield  {author} {\bibinfo {author} {\bibfnamefont {P.}~\bibnamefont
  {Czarnik}}, \bibinfo {author} {\bibfnamefont {J.}~\bibnamefont {Dziarmaga}},
  \ and\ \bibinfo {author} {\bibfnamefont {P.}~\bibnamefont {Corboz}},\
  }\bibfield  {title} {\enquote {\bibinfo {title} {Time evolution of an
  infinite projected entangled pair state: An efficient algorithm},}\ }\href
  {\doibase 10.1103/PhysRevB.99.035115} {\bibfield  {journal} {\bibinfo
  {journal} {Phys. Rev. B}\ }\textbf {\bibinfo {volume} {99}},\ \bibinfo
  {pages} {035115} (\bibinfo {year} {2019})}\BibitemShut {NoStop}%
\bibitem [{\citenamefont {Czarnik}\ and\ \citenamefont
  {Corboz}()}]{Czarnik2019b}%
  \BibitemOpen
  \bibfield  {author} {\bibinfo {author} {\bibfnamefont {P.}~\bibnamefont
  {Czarnik}}\ and\ \bibinfo {author} {\bibfnamefont {P.}~\bibnamefont
  {Corboz}},\ }\bibfield  {title} {\enquote {\bibinfo {title} {Finite
  correlation length scaling with infinite projected entangled pair states at
  finite temperature},}\ }\href@noop {} {\ }\Eprint
  {http://arxiv.org/abs/arXiv:1904.02476 (2019)} {arXiv:1904.02476 (2019)}
  \BibitemShut {NoStop}%
\bibitem [{\citenamefont {Prosen}\ and\ \citenamefont
  {Pi\ifmmode~\check{z}\else \v{z}\fi{}orn}(2007)}]{Prosen.t+:2007:Entropy}%
  \BibitemOpen
  \bibfield  {author} {\bibinfo {author} {\bibfnamefont {T.}~\bibnamefont
  {Prosen}}\ and\ \bibinfo {author} {\bibfnamefont {I.}~\bibnamefont
  {Pi\ifmmode~\check{z}\else \v{z}\fi{}orn}},\ }\bibfield  {title} {\enquote
  {\bibinfo {title} {Operator space entanglement entropy in a transverse
  {I}sing chain},}\ }\href {\doibase 10.1103/PhysRevA.76.032316} {\bibfield
  {journal} {\bibinfo  {journal} {Phys. Rev. A}\ }\textbf {\bibinfo {volume}
  {76}},\ \bibinfo {pages} {032316} (\bibinfo {year} {2007})}\BibitemShut
  {NoStop}%
\bibitem [{\citenamefont {{Barthel}}()}]{Barthel.t:2017:FiniteT}%
  \BibitemOpen
  \bibfield  {author} {\bibinfo {author} {\bibfnamefont {T.}~\bibnamefont
  {{Barthel}}},\ }\href@noop {} {\enquote {\bibinfo {title} {{One-dimensional
  quantum systems at finite temperatures can be simulated efficiently on
  classical computers}},}\ }\Eprint {http://arxiv.org/abs/arXiv:1708.09349
  (2017)} {arXiv:1708.09349 (2017)} \BibitemShut {NoStop}%
\bibitem [{\citenamefont {Dubail}()}]{Dubail17}%
  \BibitemOpen
  \bibfield  {author} {\bibinfo {author} {\bibfnamefont {J.}~\bibnamefont
  {Dubail}},\ }\bibfield  {title} {\enquote {\bibinfo {title} {Entanglement
  scaling of operators: a conformal field theory approach, with a glimpse of
  simulability of long-time dynamics in 1 + 1d},}\ }\href@noop {} {\
  }\BibitemShut {NoStop}%
\bibitem [{\citenamefont {Chen}\ \emph {et~al.}(2019)\citenamefont {Chen},
  \citenamefont {Qu}, \citenamefont {Li}, \citenamefont {Chen}, \citenamefont
  {Gong}, \citenamefont {von Delft}, \citenamefont {Weichselbaum},\ and\
  \citenamefont {Li}}]{Chen2018b}%
  \BibitemOpen
  \bibfield  {author} {\bibinfo {author} {\bibfnamefont {L.}~\bibnamefont
  {Chen}}, \bibinfo {author} {\bibfnamefont {D.-W.}\ \bibnamefont {Qu}},
  \bibinfo {author} {\bibfnamefont {H.}~\bibnamefont {Li}}, \bibinfo {author}
  {\bibfnamefont {B.-B.}\ \bibnamefont {Chen}}, \bibinfo {author}
  {\bibfnamefont {S.-S.}\ \bibnamefont {Gong}}, \bibinfo {author}
  {\bibfnamefont {J.}~\bibnamefont {von Delft}}, \bibinfo {author}
  {\bibfnamefont {A.}~\bibnamefont {Weichselbaum}}, \ and\ \bibinfo {author}
  {\bibfnamefont {W.}~\bibnamefont {Li}},\ }\bibfield  {title} {\enquote
  {\bibinfo {title} {Two-temperature scales in the triangular-lattice
  {Heisenberg} antiferromagnet},}\ }\href {\doibase 10.1103/PhysRevB.99.140404}
  {\bibfield  {journal} {\bibinfo  {journal} {Phys. Rev. B}\ }\textbf {\bibinfo
  {volume} {99}},\ \bibinfo {pages} {140404(R)} (\bibinfo {year}
  {2019})}\BibitemShut {NoStop}%
\bibitem [{\citenamefont {Mermin}\ and\ \citenamefont
  {Wagner}(1966)}]{Mermin-Wagner}%
  \BibitemOpen
  \bibfield  {author} {\bibinfo {author} {\bibfnamefont {N.~D.}\ \bibnamefont
  {Mermin}}\ and\ \bibinfo {author} {\bibfnamefont {H.}~\bibnamefont
  {Wagner}},\ }\bibfield  {title} {\enquote {\bibinfo {title} {Absence of
  ferromagnetism or antiferromagnetism in one- or two-dimensional isotropic
  {Heisenberg} models},}\ }\href {\doibase 10.1103/PhysRevLett.17.1133}
  {\bibfield  {journal} {\bibinfo  {journal} {Phys. Rev. Lett.}\ }\textbf
  {\bibinfo {volume} {17}},\ \bibinfo {pages} {1133} (\bibinfo {year}
  {1966})}\BibitemShut {NoStop}%
\bibitem [{\citenamefont {Laumann}\ \emph {et~al.}(2009)\citenamefont
  {Laumann}, \citenamefont {Parameswaran},\ and\ \citenamefont
  {Sondhi}}]{Laumann2009}%
  \BibitemOpen
  \bibfield  {author} {\bibinfo {author} {\bibfnamefont {C.~R.}\ \bibnamefont
  {Laumann}}, \bibinfo {author} {\bibfnamefont {S.~A.}\ \bibnamefont
  {Parameswaran}}, \ and\ \bibinfo {author} {\bibfnamefont {S.~L.}\
  \bibnamefont {Sondhi}},\ }\bibfield  {title} {\enquote {\bibinfo {title}
  {Absence of {Goldstone} bosons on the {Bethe} lattice},}\ }\href {\doibase
  10.1103/PhysRevB.80.144415} {\bibfield  {journal} {\bibinfo  {journal} {Phys.
  Rev. B}\ }\textbf {\bibinfo {volume} {80}},\ \bibinfo {pages} {144415}
  (\bibinfo {year} {2009})}\BibitemShut {NoStop}%
\bibitem [{\citenamefont {Lepetit}\ \emph {et~al.}(1999)\citenamefont
  {Lepetit}, \citenamefont {Cousy},\ and\ \citenamefont
  {Pastor}}]{Lepetit1999}%
  \BibitemOpen
  \bibfield  {author} {\bibinfo {author} {\bibfnamefont {M.-B.}\ \bibnamefont
  {Lepetit}}, \bibinfo {author} {\bibfnamefont {M.}~\bibnamefont {Cousy}}, \
  and\ \bibinfo {author} {\bibfnamefont {G.~M.}\ \bibnamefont {Pastor}},\
  }\bibfield  {title} {\enquote {\bibinfo {title} {Density-matrix
  renormalization study of the {Hubbard} model on a {Bethe} lattice},}\ }\href
  {https://doi.org/10.1007/s100510050053} {\bibfield  {journal} {\bibinfo
  {journal} {Eur. Phys. J. B}\ }\textbf {\bibinfo {volume} {13}},\ \bibinfo
  {pages} {421} (\bibinfo {year} {1999})}\BibitemShut {NoStop}%
\bibitem [{\citenamefont {Mahan}(2001)}]{Mahan2001}%
  \BibitemOpen
  \bibfield  {author} {\bibinfo {author} {\bibfnamefont {G.~D.}\ \bibnamefont
  {Mahan}},\ }\bibfield  {title} {\enquote {\bibinfo {title} {Energy bands of
  the {Bethe} lattice},}\ }\href {\doibase 10.1103/PhysRevB.63.155110}
  {\bibfield  {journal} {\bibinfo  {journal} {Phys. Rev. B}\ }\textbf {\bibinfo
  {volume} {63}},\ \bibinfo {pages} {155110} (\bibinfo {year}
  {2001})}\BibitemShut {NoStop}%
\bibitem [{\citenamefont {Baxter}(2013)}]{Baxter2013}%
  \BibitemOpen
  \bibfield  {author} {\bibinfo {author} {\bibfnamefont {R.~J.}\ \bibnamefont
  {Baxter}},\ }\href@noop {} {\emph {\bibinfo {title} {Exactly {{Solved
  Models}} in {{Statistical Mechanics}}}}}\ (\bibinfo  {publisher} {{Courier
  Corporation}},\ \bibinfo {year} {2013})\BibitemShut {NoStop}%
\bibitem [{\citenamefont {M{\'e}zard}\ and\ \citenamefont
  {Parisi}(2001)}]{Mezard2001}%
  \BibitemOpen
  \bibfield  {author} {\bibinfo {author} {\bibfnamefont {M.}~\bibnamefont
  {M{\'e}zard}}\ and\ \bibinfo {author} {\bibfnamefont {G.}~\bibnamefont
  {Parisi}},\ }\bibfield  {title} {\enquote {\bibinfo {title} {The {Bethe}
  lattice spin glass revisited},}\ }\href {\doibase 10.1007/PL00011099}
  {\bibfield  {journal} {\bibinfo  {journal} {Eur. Phys. J. B}\ }\textbf
  {\bibinfo {volume} {20}},\ \bibinfo {pages} {217} (\bibinfo {year}
  {2001})}\BibitemShut {NoStop}%
\end{thebibliography}%

\end{document}